\documentclass[aps,twocolumn,floatfix,groupedaddress,nofootinbib]{revtex4}
\usepackage{csquotes}
\usepackage{graphicx,color}
\usepackage{float}
\usepackage[caption=false]{subfig}
\usepackage{amsmath}
\usepackage{amsmath}
\usepackage{multirow}
\usepackage{dsfont}
\usepackage[shortlabels]{enumitem}
\usepackage{epstopdf}

\begin{document}

\title{Testing the isomorph invariance of the bridge functions of Yukawa one-component plasmas. I. Intermediate and long range}
\author{\vspace*{-2.50mm}F. Lucco Castello$^{1}$, P. Tolias$^{1}$ and J. C. Dyre$^{2}$}
\affiliation{$^{1}$Space and Plasma Physics, Royal Institute of Technology, Stockholm, SE-100 44, Sweden\\
             $^{2}$Glass and Time, IMFUFA, Roskilde University, Roskilde, DK-4000, Denmark}
\begin{abstract}
\noindent It has been recently conjectured that bridge functions remain nearly invariant along phase diagram lines of constant excess entropy for the broad class of R-simple liquids. To test this hypothesis, the bridge functions of Yukawa systems are computed outside the correlation void with the Ornstein-Zernike inversion method and structural input from ultra-accurate molecular dynamics simulations. The effect of statistical, grid, finite-size, tail and isomorphic errors is quantified. Uncertainty propagation analysis is complemented with a detailed investigation of the sensitivity of the bridge function to periodic and aperiodic multiplicative perturbations in the radial distribution function. In the long and intermediate range, bridge functions are demonstrated to be approximately isomorph invariant.
\end{abstract}
\maketitle\vspace*{-14.50mm}

\section{Introduction}\label{sec:intro}

\noindent One of the fundamental problems in the statistical mechanics of liquids concerns the accurate computation of static pair correlations with the knowledge of the pair interaction potential alone and without resorting to computer simulations. The integral equation theory of liquids features two formally exact equations for this problem that contain three unknown functions; an additional equation for the so-called bridge function is missing\,\cite{Hansenbo,hbondbok}. The unknown bridge function incorporates all the elements that make a many body problem of infinite degrees of freedom unsolvable. Therefore, it is not surprising that its diagrammatic expansion is very slowly converging and its high-order terms quickly become overly complicated to calculate\,\cite{bridgev1}. As a consequence, numerous approximation schemes have been developed for the bridge function, whose effectiveness varies depending on the potential and can only be reliably evaluated a posteriori through a comparison with \enquote{exact} simulation results\,\cite{Hansenbo,bridgev3,bridgev4}. It is fortuitous, though, that static correlations in liquids exhibit relatively weak dependence on the bridge function\,\cite{Hansenbo}.

Contrary to the radial distribution function, the bridge function neither possesses microscopic representation (in terms of $\delta-$functions and the instantaneous particle positions) nor a conditional probability interpretation. Thus, the bridge function cannot be directly extracted from computer simulations. Nevertheless, it can be computed with input from computer simulations. In particular, extraction of radial distribution functions leads to the closure of the system of integral equation theory and allows one to solve for the unknown bridge function. There are two main caveats with such an inversion method. First, the weak dependence of static correlations on the bridge function implies a strong sensitivity of the bridge function on the radial distribution function, in other words what is an asset in the direct problem becomes an obstacle in the inverse problem, which necessitates long simulations with large particle number. Second, irrespective of the achieved size of the statistical sample, the inversion method is doomed to fail at very short distances where the probability of encountering another particle is ultra low (for finite thermodynamically stable potentials) or is even zero (for unbounded potentials). Cavity simulations featuring a special tagged particle pair and utilizing umbrella sampling techniques have been conceived for the computation of the bridge function at such distances\,\cite{bridgHSO}.

In spite of the inherent numerical difficulties of the computational procedure, \enquote{exact} bridge functions have been computed with structural input from Monte Carlo (MC) or molecular dynamics (MD) simulations for different established model potentials such as hard sphere systems\,\cite{bridgHSO,bridgHS0,bridgHS1,bridgHS2} as well as their binary mixtures\,\cite{bridgHS0,bridgHS2}, Lennard-Jones systems\,\cite{bridgLJ1,bridgLJ2,bridgLJ3,bridgLJ4}, inverse power law systems\,\cite{bridIPL1,bridIPL2} and one-component plasma liquids\,\cite{PollOCPB,bridOCP1,bridOCP2}. In addition, \enquote{exact} bridge functions have been obtained for more realistic liquids including isotropic hard spheroid fluids\,\cite{bridgHS3}, liquid metal model inter-ionic potentials\,\cite{bridgme1,bridgme2}, model 2-2 electrolytes\,\cite{bridgele}, molten salts\,\cite{bridgsal}, Lennard-Jones dipolar fluids\,\cite{bridgSTO} and even the extended simple point charge (SPC/E) site-site model of water\,\cite{bridgwat}.

Despite the undeniable progress during half a century of investigations, few exact or approximate properties of bridge functions have so far been discovered. Recently, a novel integral equation theory approach has been formulated that is based on the conjecture that bridge functions remain invariant along phase diagram lines of constant excess entropy for a broad class of liquids known as R-simple\,\cite{ourwork1}. It has been coined as isomorph-based empirically modified hypernetted-chain (IEMHNC) and has been applied to Yukawa and bi-Yukawa liquids resulting in a remarkable agreement with simulations\,\cite{ourwork1,ourwork2,ourwork3}.

The main objective of the present study is to test the validity of the underlying ansatz of bridge function invariance for Yukawa one-component plasmas. In the present article, the intermediate and long range of Yukawa bridge functions will be computed along isentropic lines with ultra-accurate MD simulations. In the accompanying article, the short range of Yukawa bridge functions will be computed with specially designed MD simulations\footnotemark\footnotetext{To our knowledge, Yukawa bridge functions have not been computed before from simulations. Screening potentials have been computed from MC simulations\,\cite{bridgYuk}, which strongly differ from bridge functions not requiring the direct correlation functions.}.

The paper is organized as follows. Section \ref{Background} features an introduction to Yukawa one-component plasmas, isomorph theory, R-simple systems and discusses arguments in favor of the isomorph invariance of bridge functions. In section \ref{Isomorphs}, $16$ Yukawa state points are identified with isomorph tracing techniques distributed amongst four isomorphs. In section \ref{Inversion}, the inversion method is presented, MD simulation parameters are specified and bridge functions are computed for all state points. In section \ref{SyntheticUncertainties}, the bridge function sensitivity to artificial periodic \& aperiodic multiplicative perturbations of radial distribution functions is studied. In section\,\ref{RealUncertainties}, the effect of statistical, grid, finite size, tail and isomorphic errors in the bridge function is quantified. In section \ref{InversionErrors}, corrected bridge functions with error bars due to propagating uncertainties are presented and the degree of isomorph invariance is discussed. In section\,\ref{Summary}, the results are summarized.

\section{Background}\label{Background}

\noindent For the present article to be self-contained, following a brief introduction to the standard nomenclature of the Yukawa one-component plasmas, a concise primer focusing on the isomorph theory of R-simple systems and the isomorph-based empirically modified hypernetted-chain approach is given in the present section. The reader is addressed to the references cited below for further details.

\subsection{Yukawa one-component plasmas}\label{subYOCP}

\noindent \emph{Yukawa one-component plasma (YOCP)} systems are model systems whose constituents are charged point particles which are immersed in a neutralizing background and interact via the screened Coulomb (Yukawa) pair potential $u(r)=(Q^2/r)\exp(-r/\lambda)$ where $Q$ is the particle charge and $\lambda$ the screening length determined by the polarizable background. It is convenient to specify the thermodynamic state points of the YOCP in terms of two independent dimensionless variables, the coupling parameter $\Gamma$ and the screening parameter $\kappa$ defined by\,\cite{dustrev1,dustrev2,dustrev3,dustrev4}
\begin{equation*}
\Gamma=\beta\frac{Q^2}{d}\,,\qquad\kappa=\frac{d}{\lambda}\,.
\end{equation*}
In the above; $\beta=1/(k_{\mathrm{B}}T)$ with $k_{\mathrm{B}}$ the Boltzmann constant, $T$ the temperature and $d=(4\pi{n}/3)^{-1/3}$ for the Wigner-Seitz radius with\,$n$\,the particle (number) density.

In the limit of a rigid background $\lambda\to\infty$ or $\kappa\to0$, the Yukawa potential collapses to the unscreened Coulomb potential and the resulting model system is then known as one-component plasma (OCP). The YOCP enables the exploration of the full range of potential softness from the long range Coulomb interactions of the OCP for $\kappa=0$ to ultra-short range hard-sphere interactions for $\kappa\to\infty$. Due to its variable softness and its relevance to strongly coupled laboratory systems such as complex plasmas and colloidal suspensions\,\cite{dustrev6,dustrev7}, the YOCP is still being actively investigated in statistical mechanics studies.

It is worth noting that the distances are typically normalized by the Wigner-Seitz radius $d=(4\pi{n}/3)^{-1/3}$ in the non-ideal plasma literature, while the distances are typically normalized by the mean-cubic inter-particle distance $\Delta=n^{-1/3}$ in the liquid state and isomorph theory literature. Both normalizations will be used in the present work, but mainly the plasma normalization to remain consistent with the screening parameter definition.

\subsection{Isomorph theory and R-simple systems}\label{subIsomorphRsimple}

\noindent \emph{Isomorphic lines} or simply \emph{isomorphs} are phase diagram curves of constant excess entropy, along which a large set of structural and dynamic properties are approximately invariant when expressed in properly reduced units\,\cite{isogene1,isogene2,isogene3}. In case of Newtonian dynamics, the length is normalized to the mean-cubic inter-particle distance $\Delta=n^{-1/3}$, the energy is normalized to the thermal energy $k_{\mathrm{B}}T$ and the time is normalized to the time required for a particle that is free streaming with its thermal velocity to traverse an inter-particle distance $\tau=n^{-1/3}\sqrt{m/(k_{\mathrm{B}}T)}$\,\cite{isogene3}. All systems have isentropic curves in their thermodynamic phase diagram, but these are termed isomorphs only for the so-called Roskilde-simple or R-simple systems.

\emph{R-simple systems are rigorously defined} as many body systems that possess the property that the ordering of the total potential energies of two configurations consistent with the same density is maintained when these two configurations are uniformly scaled to a different density\,\cite{isogene4}. Mathematically, $U(\boldsymbol{R}_{\mathrm{a}})<U(\boldsymbol{R}_{\mathrm{b}})\Rightarrow{U}(\mu\boldsymbol{R}_{\mathrm{a}})<U(\mu\boldsymbol{R}_{\mathrm{b}})$ for positive $\mu$, where $U(\boldsymbol{R})$ is the total potential energy, $\boldsymbol{R}$ is the particle configuration that is given by the collective N-particle position vector $(\boldsymbol{r}_1,...,\boldsymbol{r}_N)$ and with $\boldsymbol{R}_{\mathrm{a}},\,\boldsymbol{R}_{\mathrm{b}}$ denoting two equal density configurations\,\cite{isogene4}. The hidden scale invariance property is exact only for systems that are characterized by Euler-homogeneous interactions (plus a constant), such as inverse power law (IPL) systems. For other R-simple systems, hidden scale invariance should be understood to be valid for most of the physically relevant configurations reflecting the approximate nature of isomorph theory.

\emph{R-simple systems are practically identified} as systems possessing strong correlations between their virial $(W)$ \& potential energy $(U)$ constant-volume thermal equilibrium fluctuations\,\cite{isogene5}. The degree of $W-U$\,correlations is quantified by the standard Pearson coefficient given by
\begin{equation*}
R_{WU}=\frac{\langle\Delta{W}\Delta{U}\rangle_{\mathrm{NVT}}}{\langle(\Delta{W})^2\rangle_{\mathrm{NVT}}\langle(\Delta{U})^2\rangle_{\mathrm{NVT}}}\,,
\end{equation*}
where $\langle...\rangle_{\mathrm{NVT}}$ denotes canonical ensemble averaging and $\Delta{A}=A-\langle{A}\rangle_{\mathrm{NVT}}$ denotes statistical fluctuations around the canonical mean. Strong $W-U$ correlations are empirically delimited by the practical condition $R_{WU}\gtrsim0.9$ that allows for straightforward identification of R-simple systems with canonical (NVT) computer simulations.

A recent computational investigation revealed that the YOCP is an R-simple system that exhibits exceptionally strong $W-U$ correlations ($R_{WU}>0.99$) for an extended part of the fluid phase covering the entire dense liquid region of the phase diagram\,\cite{isoYOCPg}. This rationalizes a number of previous observations such as the fact that the YOCP excess internal energies conform to the Rosenfeld-Tarazona decomposition\,\cite{isogene6,isogene7} as well as the fact that the YOCP reduced transport coefficients strongly abide to Rosenfeld's excess entropy scaling\,\cite{isogene8,isogene9}.

\subsection{The isomorph-based empirically modified hypernetted chain approximation}\label{subIEMHNC}

\noindent The \emph{isomorph-based empirically modified hypernetted chain (IEMHNC) approximation} is an integral equation theory approach that is based on the assumption of isomorph invariance of bridge functions when expressed in reduced distance units\,\cite{ourwork1}. The invariance ansatz closes the non-linear non-local equation system that arises in integral equation theory provided that two external inputs are also available: a closed-form expression for the dependence of the isomorphic curves on the thermodynamic state points and a closed-form bridge function expression that is valid along any phase diagram line that possesses a unique intersection point with any isomorphic curve\,\cite{ourwork1}. With such input, the isentropic correspondence maps the bridge function from the initial phase diagram line to the entire phase diagram.

The IEMHNC approach has been successfully applied to dense Yukawa and bi-Yukawa liquids\,\cite{ourwork1,ourwork2,ourwork3} taking advantage of an established parameterization of the OCP bridge function through the reduced distance and coupling parameter\,\cite{bridOCP1}. Comprehensive benchmarking with computer simulations has revealed that the IEMHNC approach possesses a remarkable accuracy with predictions of structural properties within $2\%$ inside the first coordination cell and predictions of thermodynamic properties within $0.5\%$ in the entire dense liquid regime\,\cite{ourwork1,ourwork2}. In addition, systematic comparison with different advanced integral equation theory approaches has demonstrated that the performance of the IEMHNC approach is comparable to that of the variational modified hypernetted-chain approach (VMHNC)\,\cite{rosVMHNC} but with $10-80$ times less computational cost depending on the state point\,\cite{ourwork3}.

\subsection{Theoretical arguments in favor of the isomorph invariance of bridge functions}\label{subArguments}

\noindent The excellent performance of the IEMHNC approach for YOCP systems\,\cite{ourwork1} and for biYOCP systems\,\cite{ourwork2} clearly suggests that the underlying conjecture of the isomorph invariance of bridge functions must hold to a high degree. This is also indicated by the fact that the IEMHNC approach preserves its OCP level of accuracy regardless of the value of the YOCP or biYOCP screening parameter. Moreover, the VMHNC bridge function has been revealed to be implicitly isomorph invariant for the YOCP, since the effective packing fraction acquired by minimizing the respective free energy functional has been demonstrated to remain nearly constant along any isomorphic curve within the dense liquid regime\,\cite{ourwork3}. Finally, the output of the classic hypernetted-chain (HNC) approach, that completely neglects all the bridge diagrams, leads to approximately invariant static correlations for the YOCP\,\cite{ourwork1}, which implies that the addition of an isomorph invariant bridge function would be beneficial for this isomorph invariance to persist.

Further arguments in support of the bridge function isomorph invariance are connected to the notion of bridge function quasi-universality\,\cite{invmeth1} that forms the backbone of the powerful modified hypernetted-chain (MHNC) and reference hypernetted-chain (RHNC) approaches. This quasi-universality notion can be summarized in the statement that, in their short range, the bridge functions constitute the same universal family of curves irrespective of the interaction potential and it was based on the fact that bridge functions can be expressed as densely connected diagrams containing total correlation function bonds\,\cite{invmeth1}. Considering the isomorph invariance of the total correlation functions, the same reasoning can be extended to the notion of isomorph invariance. The isomorph theory has already rationalized a number of well-established quasi-universalities of simple liquids, since the excess entropy always turned out to be the controlling parameter\,\cite{isogene3}. In addition, the isomorph invariance of bridge functions is consistent with the zero-separation bridge function freezing criterion of Rosenfeld, which states that the value of the bridge function at the origin $r=0$, when calculated along the liquid-solid phase transition line, is nearly constant and even independent of the pair potential\,\cite{zerosep1}. This criterion is known to be satisfied for the YOCP\,\cite{zerosep2}.

Finally, let us discuss an apparent incompatibility of bridge function isomorph invariance with the condition of unique functionality. This condition assumes that the exact functional relation between the bridge function and indirect correlation function approximately reduces to a unique function\,\cite{uniqfun1}. This condition is implicitly invoked in most fundamental bridge function closures of integral equation theory\,\cite{uniqfun1}. However, given the isomorph variance of the indirect correlation function (to be revealed in the following sections), it also implies that the bridge function cannot be isomorph invariant. Conversely, the approximate bridge function properties of isomorph invariance and unique functionality are incompatible. This does not constitute a contradiction, since it has been revealed (with the use of Duh-Haymet plots) that the above formulation of the unique functionality condition does not hold\,\cite{uniqfun2}. In fact, the optimized unique functionality conditions that are invoked in more modern approaches feature a re-normalized indirect correlation function, typically stemming from a partition of the interaction potential\,\cite{uniqfun3,uniqfun4,uniqfun5,uniqfun6}. Such formulations are not incompatible with the isomorph invariance ansatz.

\section{Isomorph tracing and state points of interest}\label{Isomorphs}

\noindent Different methods are available for the tracing of the isomorph curves of R-simple systems with or even without the use of computer simulations. In the present investigation, we shall employ the direct isomorph check, the small step method and the analytical method. The physical basis and the numerical implementation of these methods are briefly described below.

The \emph{direct isomorph check} is based on an approximate relation valid for any state point that is a fundamental characteristic of R-simple systems and reads as\,\cite{isogene4}
\begin{equation*}
U(\boldsymbol{R})=U[n,S_{\mathrm{ex}}(\widetilde{\boldsymbol{R}})]\,,
\end{equation*}
where $U(\boldsymbol{R})$ is the instantaneous potential energy which depends on the configuration $\boldsymbol{R}$ consistent with any state point $(n,T)$, $S_{\mathrm{ex}}(\widetilde{\boldsymbol{R}})$ is the instantaneous excess entropy function that depends on the reduced configuration $\widetilde{\boldsymbol{R}}=n^{1/3}\boldsymbol{R}$ and $U(n,S_{\mathrm{ex}})$ is the thermodynamic (ensemble averaged) potential energy. Let us suppose a $(n_1,T_1)$ reference state together with its isomorphic $(n_2,T_2)$ state of re-scaled density $n_2=(1/\mu^3)n_1$ but unknown temperature $T_2$. Let us also consider the configurations $\boldsymbol{R}_1$, $\boldsymbol{R}_2$ of these state points that are identical in reduced units, $n_1^{1/3}\boldsymbol{R}_1=n_2^{1/3}\boldsymbol{R}_2$. Application of the above relation for both the state points, first-order Taylor expansion with respect to $S_{\mathrm{ex}}(\widetilde{\boldsymbol{R}})$ around the thermodynamic excess entropy $S_{\mathrm{ex}}$, utilization of the identity $(\partial{U}/\partial{S}_{\mathrm{ex}})_{n}=T$ as well as use of the identical reduced entropies and reduced configurations, leads to the approximate expression\,\cite{isogene4}
\begin{equation*}
\frac{U(\boldsymbol{R}_1)-U_1}{U(\boldsymbol{R}_2)-U_2}\simeq\frac{T_1}{T_2}\,.
\end{equation*}
The above expression constitutes the basis of the direct isomorph check that in practice works as follows\,\cite{isomdic2,isomdic3}; The potential energy $U(\boldsymbol{R}_1)$ is extracted from a $(n_1,T_1)$ simulation, the configuration is rescaled to $\boldsymbol{R}_2=\mu\boldsymbol{R}_1$ and the potential energy $U(\boldsymbol{R}_2)$ is extracted. Repetition of this procedure for numerous $\boldsymbol{R}_1$ configurations leads to a scatter plot between $U(\boldsymbol{R}_1)$ and $U(\boldsymbol{R}_2)$ that is well approximated by a straight line and whose linear regression slope $T_1/T_2$ allows the determination of the unknown $T_2$.

In the present application of the direct isomorph check, the algorithm is formulated in terms of $(\Gamma,\kappa)$ and a fixed screening parameter jump $\Delta\kappa/\kappa=3.1\%$ is considered which translates to a $|\Delta{n}|/n=9.8\%$ density variation between successive isomorphic state points. In the NVT MD simulations that are necessary for the slope extraction, reduced units are employed by setting the temperature and density equal to unity and controlling the length and energy parameters of the potential. The interaction potential is truncated at $r_{\mathrm{cut}}=10\Delta$ with the shifted-force cutoff method, the time step is $\Delta{t}/\tau=2.5\times10^{-3}$, the equilibration time is $2^{20}\Delta{t}$, the statistics duration is $2^{20}\Delta{t}$, the saving period is $2^{10}\Delta{t}$, and the number of particles is $8192$ ($20\Delta$ for the simulation box length).

The \emph{small step method} combines the thermodynamic definition of the so-called density-scaling exponent with an exact alternative expression that originates from thermodynamic fluctuation theory. The density-scaling exponent $\gamma(n,T)$ is defined in log-log density-temperature phase diagrams as the local slope of the isentropic line traversing the state point $(n,T)$. Thus, we have\,\cite{isomssm1,isomssm2}
\begin{equation*}
\gamma(n,T)=\left(\frac{\partial\ln{T}}{\partial\ln{n}}\right)_{S_{\mathrm{ex}}},
\end{equation*}
with $S_{\mathrm{ex}}$ denoting the excess entropy. The density-scaling exponent is also acquired by the linear regression slope of the scatter plot between the virial and potential energy canonical fluctuations, since we also have\,\cite{isomssm1,isomssm2}
\begin{equation*}
\gamma(n,T)=\frac{\langle\Delta{U}\Delta{W}\rangle_{\mathrm{NVT}}}{\langle(\Delta{U})^2\rangle_{\mathrm{NVT}}}\,.
\end{equation*}
The fluctuation theory expression allows the evaluation of $\gamma(n,T)$ at any state point from canonical simulations, whereas the thermodynamic definition constitutes an explicit non-linear first-order differential equation with respect to the state points that can be solved with any numerical scheme in order to trace the respective isomorph. This procedure has been coined as small step method, because its typical applications utilize first-order numerical schemes for the solution of the differential equation that necessitate small steps in the density\,\cite{isomssm3,isomssm4}. However, implementation of higher-order schemes, such as the classical fourth-order Runge-Kutta method (RK4), allows for larger density increments and leads to equally accurate isomorph tracing but with far less computational cost.

In the present application of the small step method, the RK4 algorithm is formulated in terms of $(\ln{n},\,\ln{T})$ and a fixed logarithmic density step is considered which translates to a $|\Delta{n}|/n=8.8\%$ density variation between successive isomorphic state points. In the NVT MD simulations that are necessary for $\gamma$ extraction, natural units $(n,T)$ are employed, the Yukawa pair potential is truncated at $r_{\mathrm{cut}}=10d$ with the shifted-force cutoff method, the time step is $\Delta{t}/\tau=2.5\times10^{-3}$, the equilibration time is $2^{17}\Delta{t}$, the statistics duration is $2^{17}\Delta{t}$, the saving period is $2^{7}\Delta{t}$, and the number of particles is $8192$ ($32d$ for the simulation box length).

The \emph{analytical method} exploits a number of exact properties of inverse power law potentials in order to define an approximate distance dependent IPL-like exponent for arbitrary pair potentials\,\cite{isomana1}. Complemented with a realistic estimate for the effective nearest-neighbor distance, such a definition leads to an approximate relation for the density-scaling exponent and a closed-form expression for any family of isomorphic curves\,\cite{isoYOCPg,isomana1}. The application of the analytical method to YOCP systems results in\,\cite{isoYOCPg}
\begin{equation*}
\Gamma_{\mathrm{ISO}}(\kappa){e}^{-\Lambda\alpha\kappa}\left[1+\Lambda\alpha\kappa+\frac{1}{2}(\Lambda\alpha\kappa)^2\right]=\mathrm{const.}\,,
\end{equation*}
where $\Lambda$ denotes a weakly state-point dependent parameter with a value close to unity and $\alpha=\Delta/d=(4\pi/3)^{1/3}$ denotes the ratio between the mean-cubic inter-particle distance and the Wigner-Seitz radius. The simple choice $\Lambda=1$ has proven to be very accurate for the YOCP\,\cite{isoYOCPg},
\begin{equation}
\Gamma_{\mathrm{ISO}}(\kappa){e}^{-\alpha\kappa}\left[1+\alpha\kappa+\frac{1}{2}(\alpha\kappa)^2\right]=\mathrm{const.}\label{analytical_isomorph}
\end{equation}
The above expression is identical to the well-known semi-empirical description of the YOCP melting line\,\cite{isomana2,isomana3}
\begin{equation}
\Gamma_{\mathrm{m}}(\kappa){e}^{-\alpha\kappa}\left[1+\alpha\kappa+\frac{1}{2}(\alpha\kappa)^2\right]=\Gamma_{\mathrm{m}}^{\mathrm{OCP}}\,,\label{analytical_melting}
\end{equation}
which accurately follows the near-exact data obtained by MD simulations\,\cite{isomana4,isomana5}. In the above, $\Gamma_{\mathrm{m}}^{\mathrm{OCP}}=171.8$ is the OCP coupling parameter at melting\,\cite{isomana4}. It is evident that all isomorph lines are nearly parallel to the melting line, an observation that is true for any R-simple system to the first order\,\cite{isomssm1,isomana6}.

\begin{table}
	\caption{The $\kappa=(1.0,\,1.5,\,2.0,\,2.5)$ members of the $\Gamma_{\mathrm{ISO}}^{\mathrm{OCP}}=(160,120,80,40)$ or $\Gamma/\Gamma_{\mathrm{m}}=(0.93,0.70,0.47,0.23)$ isomorphs. The coupling parameters resulting from the analytical method ($\Gamma_{\mathrm{ana}}$), direct isomorph check ($\Gamma_{\mathrm{dic}}$) and small step method ($\Gamma_{\mathrm{ssm}}$) are reported. The absolute relative deviations between $\Gamma_{\mathrm{ana}}$ \& $\Gamma_{\mathrm{dic}}$ are denoted with $e_{\mathrm{ana/dic}}$, whereas those between $\Gamma_{\mathrm{ssm}}$ \& $\Gamma_{\mathrm{dic}}$ are denoted with $e_{\mathrm{ssm/dic}}$. The correlation coefficient between virial and potential energy fluctuations $R_{WU}$ is reported together with the density scaling exponent $\gamma$.}\label{isomorphtable}
	\centering
	\begin{tabular}{cccccccc}\hline
$\kappa$ & $\Gamma_{\mathrm{ana}}$ & $\Gamma_{\mathrm{dic}}$ & $\Gamma_{\mathrm{ssm}}$ & $e_{\mathrm{ana/dic}}$ & $e_{\mathrm{ssm/dic}}$ & $R_{WU}$  & $\gamma$ \\ \hline
1.0\,    &  \,205.061\,            &  \,205.061\,            &  \,205.061\,            & 0.00\%      		    &   0.00\%               & \,0.990\, & \,0.522  \\
1.5\,    &  \,283.178\,            &  \,286.437\,            &  \,286.289\,            & 1.14\%  			    &	0.05\%               & \,0.990\, & \,0.712  \\
2.0\,    &  \,426.757\,            &  \,435.572\,            &  \,435.268\,            & 2.02\%  			    &	0.07\%               & \,0.994\, & \,0.939  \\
2.5\,    &  \,684.511\,            &  \,708.517\,            &  \,707.487\,            & 3.39\%  			    &	0.14\%               & \,0.996\, & \,1.192  \\
1.0\,    &  \,153.796\,            &  \,153.796\,            &  \,153.796\,            & 0.00\%      			&	0.00\%               & \,0.991\, & \,0.523  \\
1.5\,    &  \,212.383\ ,           &  \,215.930\,            &  \,215.542\,            & 1.64\%  			    &	0.18\%               & \,0.992\, & \,0.715  \\
2.0\,    &  \,320.068\,            &  \,328.816\,            &  \,328.710\,            & 2.66\%  			    &	0.03\%               & \,0.994\, & \,0.938  \\
2.5\,    &  \,513.384\,            &  \,534.722\,            &  \,534.034\,            & 3.99\%  			    &	0.13\%               & \,0.995\, & \,1.189  \\
1.0\,    &  \,102.531\,            &  \,102.531\,            &  \,102.531\,            & 0.00\%      			&	0.00\%               & \,0.988\, & \,0.529  \\
1.5\,    &  \,141.589\,            &  \,144.330\,            &  \,144.325\,            & 1.90\%  			    &	0.00\%               & \,0.991\, & \,0.720  \\
2.0\,    &  \,213.378\,            &  \,219.972\,            &  \,220.496\,            & 3.00\%  			    &	0.24\%               & \,0.992\, & \,0.940  \\
2.5\,    &  \,342.256\,            &  \,357.136\,            &  \,358.051\,            & 4.17\%  			    &	0.26\%               & \,0.995\, & \,1.178  \\
1.0\,    &  \,51.265\,             &  \,51.265\,             &  \,51.265\,             & 0.00\%      			&	0.00\%               & \,0.988\, & \,0.534  \\
1.5\,    &  \,70.794\,             &  \,72.537\,             &  \,72.748\,             & 2.40\%  			    &	0.29\%               & \,0.988\, & \,0.730  \\
2.0\,    &  \,106.689\,            &  \,110.707\,            &  \,111.094\,            & 3.63\%  			    &	0.35\%               & \,0.990\, & \,0.934  \\
2.5\,    &  \,171.128\,            &  \,178.269\,            &  \,179.217\,            & 4.01\% 	            &	0.53\%               & \,0.993\, & \,1.166  \\ \hline \hline
	\end{tabular}
\end{table}

In the present work, the analytical method will only be invoked in order to specify the OCP members, $\Gamma_{\mathrm{ISO}}^{\mathrm{OCP}}$, of YOCP isomorphs. The mapping should be very accurate, since Eqs.(\ref{analytical_isomorph},\ref{analytical_melting}) are nearly exact for $\kappa\lesssim1.5$. In addition, the equivalence of Eqs.(\ref{analytical_isomorph},\ref{analytical_melting}) results in $\Gamma/\Gamma_{\mathrm{m}}=\mathrm{const.}\leq1$ along any distinct isomorph. For brevity, in what follows, the constant approximate values of $\Gamma_{\mathrm{ISO}}^{\mathrm{OCP}}$ or $\Gamma/\Gamma_{\mathrm{m}}$ will be utilized in order to uniquely identify the isomorphs.

\begin{figure}
	\centering
	\includegraphics[width=3.4in]{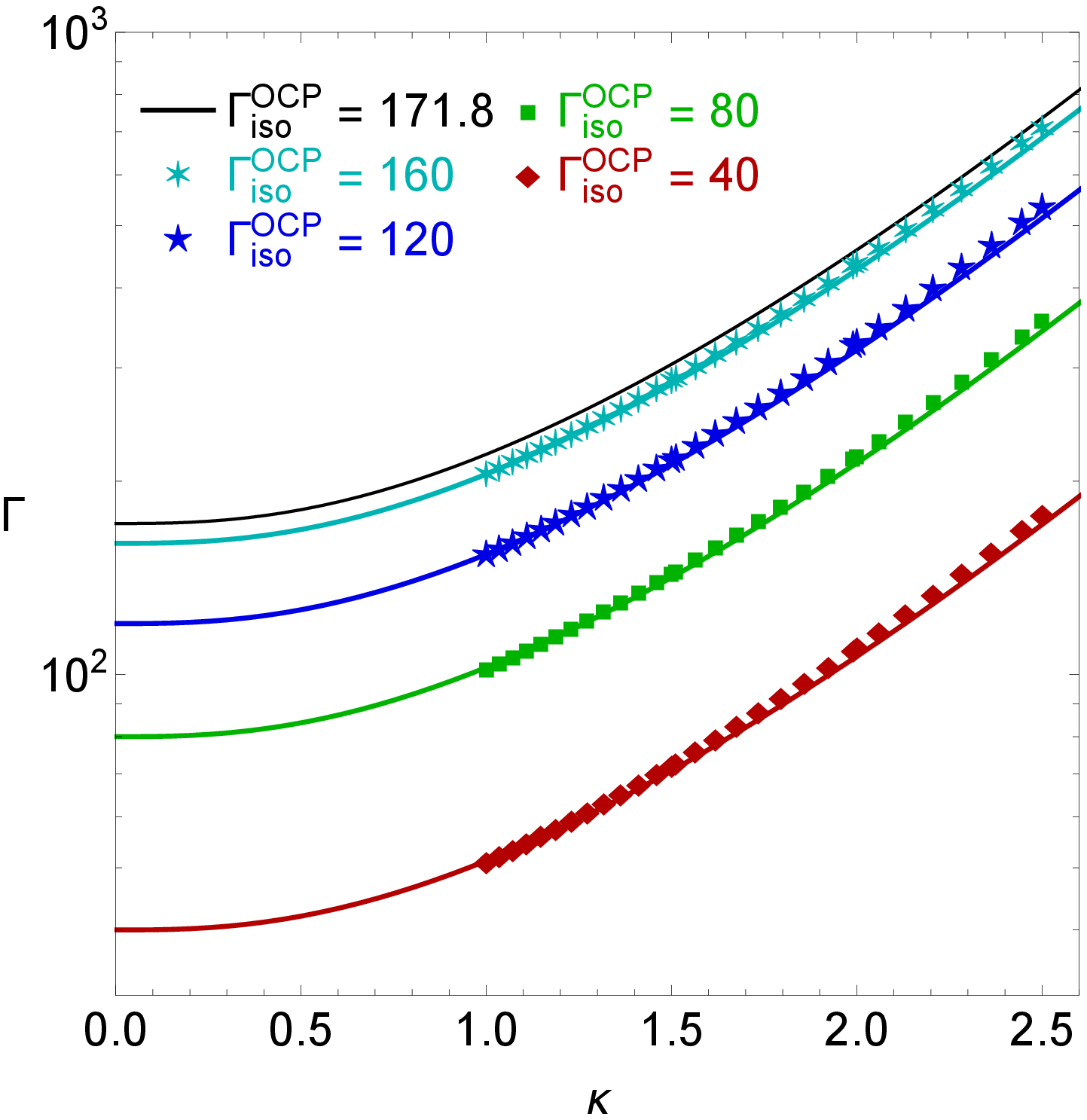}
	\caption{The $4$ targeted isomorphs $\Gamma_{\mathrm{ISO}}^{\mathrm{OCP}}=(160,120,80,40)$ together with the approximate melting line $\Gamma_{\mathrm{ISO}}^{\mathrm{OCP}}=171.8$ for dense YOCP liquids in the $\log{\Gamma}-\kappa$ phase diagram. The four isomorphic curves as obtained from the direct
    isomorph check (discrete symbols) are compared to those obtained from the approximate analytical expression (solid lines). The four numerical isomorphs begin to overshoot the respective analytical isomorphs roughly above $\kappa\simeq1.5$, see also Table \ref{isomorphtable}.}\label{fig:phase_diagram}
\end{figure}

YOCP bridge functions will be determined for $16$ state points that are equally spread amongst four different isomorphic curves. The \emph{normalized screening parameters} $\kappa$ of interest are $\kappa=(1.0,\,1.5,\,2.0,\,2.5)$, since for $\kappa\lesssim1$ the YOCP behavior becomes nearly OCP like while for $\kappa\gtrsim3$ the YOCP behavior becomes nearly hard sphere like. Such values are typically realized in complex plasma microgravity experiments\,\cite{dustexp1,dustexp2}. The \emph{OCP coupling parameters} of interest are $\Gamma_{\mathrm{ISO}}^{\mathrm{OCP}}=(160,\,120,\,80,\,40)$ and span the whole dense YOCP liquid regime, since they correspond to $\Gamma/\Gamma_{\mathrm{m}}=(0.93,\,0.70,\,0.47,\,0.23)$, respectively. The $\kappa=1$ members of these four isomorphs are then calculated with the analytical method, Eq.(\ref{analytical_isomorph}), that leads to $\Gamma_{\mathrm{ISO}}^{\kappa=1}=(205.061,\,153.796,\,102.531,\,51.265)$ which are the starting state points of the direct isomorph check and the small step method. The isomorphic coupling parameters are determined exactly at the remaining screening parameters ($\kappa=1.5,\,2.0,\,2.5$) by a targeted jump from the closest $\kappa$ point that emerges from the algorithms depending on their assumed density variations.

The YOCP state points of interest are listed in Table \ref{isomorphtable}. In all cases, the relative deviations between the results of the direct isomorph check and the small step method are always less than $0.53\%$. Since the two MD implementations are characterized by the same number of particles ($2^{13}$) and statistically useful configurations ($2^{10}$), this should be a consequence of the very high virial - potential energy fluctuations $R_{WU}\geq0.988$. This near-unity correlations imply that the starting equation of the direct isomorph check is exact, similar to the starting equation of the small step method. The YOCP state points that stem from the direct isomorph check were selected for bridge function computation. On the other hand, the absolute relative deviations between the results of the analytical method and the direct isomorph check can reach $4.17\%$. A consistent overestimation is observed for the analytical method above $\kappa\simeq1.5$, as better illustrated in figure \ref{fig:phase_diagram}. This is rather expected because the analytical expression for the melting line, see Eq.(\ref{analytical_melting}), also overshoots the MD results above $\kappa\simeq1.5$. Notice that the density-scaling exponent $\gamma$, constant for inverse power law systems, varies from $0.5\,(\kappa=1.0)$ to $1.2\,(\kappa=2.5)$ within each isomorph.

Finally, it is worth pointing out that the above isomorph tracing methods manage to identify the isentropic points of R-simple systems without ever calculating the excess entropy. Accurate theoretical determination of the excess entropy can be formidable due to the need for either thermodynamic integration or high-order correlation inclusion (see the Nettleton-Green expansion)\,\cite{NetGreen}. The same applies for the computational determination owing to the inefficiency of Widom test particle insertion methods at high densities\,\cite{isogene9}. For completeness, an estimate of the reduced excess entropy of each isomorph line has been attempted based on the equation of state suggested by Hamaguchi \emph{et al.}\,\cite{isomana4,isomana5}. This led to $(\Gamma_{\mathrm{ISO}}^{\mathrm{OCP}},-s_{\mathrm{ex}})=\{(40,2.02),\,(80,2.87),\,(120,3.49),\,(160,3.99)\}$.

\section{Bridge functions determined by the Ornstein-Zernike inversion method}\label{Inversion}

\noindent By inspecting the building blocks of the integral equation theory of liquids, it becomes evident that the bridge function acts as an additional many-body component of the pair interaction potential\,\cite{invmeth1}. Hence, the computational technique utilized for the deduction of bridge functions from simulation structural data is identical to the computational method employed for the deduction of interaction potentials from experimental structural data\,\cite{invmeth2,invmeth3}. We will refer to it as Ornstein-Zernike inversion method, since the aforementioned determination of the pair interaction potential is known as the inverse problem\,\cite{invmeth4}. In this section, the inversion method will be described, the parameters of the production runs or test simulations will be provided and the numerical results will be analyzed.

\subsection{The computational method}\label{subComputationalMethod}

\noindent In the case of one-component pair-interacting isotropic systems, the integral equation theory of liquids consists of the convolution-type Ornstein-Zernike (OZ) equation\,\cite{Hansenbo}
\begin{equation}
h(r)=c(r)+n\int c(r')h(|\boldsymbol{r}-\boldsymbol{r}'|)d^3r'\,,\label{eq:theory_oz}
\end{equation}
combined with the following exact non-linear closure condition that is derived from cluster diagram analysis\,\cite{Hansenbo}
\begin{equation}
g(r)=\exp\left[-\beta u(r)+h(r)-c(r)+B(r)\right]\,,\label{eq:theory_oz_closure}
\end{equation}
where $g(r)$ is the radial distribution function, $c(r)$ is the direct correlation function, $h(r)=g(r)-1$ is the total correlation function and $B(r)$ the bridge function. Other static correlation functions of relevance concern the indirect correlation function $\gamma(r)=h(r)-c(r)$, the potential of mean force $\beta{w}(r)=-\ln{[g(r)]}$ and the screening potential $\beta{H}(r)=\beta{u}(r)-\beta{w}(r)$. A formally exact expression for the bridge function that is required to close the system of equations is unavailable due to the highly connected nature of bridge diagrams. In addition, the exact virial-type series that define bridge functions through Mayer functions or total correlation functions converge very slowly and are notoriously hard to compute\,\cite{bridgev1,bridgev2}. For this reason, integral equation theory approaches invoke approximations that either prescribe $B(\gamma)$ of a function of the indirect correlation function or parameterize $B(r/d)$ as a function of the normalized distance\,\cite{Hansenbo,bridgev3,bridgev4}.

In integral equation theory approaches, Eqs.(\ref{eq:theory_oz},\ref{eq:theory_oz_closure}) are solved for $[g(r),c(r)]$ with known $u(r)$ and an assumption for $B(r)$ which suggests that the equations are coupled. In pair interaction reconstruction, Eqs.(\ref{eq:theory_oz},\ref{eq:theory_oz_closure}) are solved for $[c(r),u(r)]$ with known $g(r)$ and an assumption for $B(r)$ which suggests that multiple viable solutions can emerge. In bridge function reconstruction, Eqs.(\ref{eq:theory_oz},\ref{eq:theory_oz_closure}) are solved for $[c(r),B(r)]$ with known $[g(r),u(r)]$ which suggests that the equations are decoupled and a unique solution exists.

The computation of bridge functions from MD simulations proceeds in the following manner. \textbf{(1)} The radial distribution function is extracted from MD simulations with the histogram method. Constant $\Delta/r=0.002d$ bin widths are assumed, see section \ref{subErrorGrid} for a detailed justification. \textbf{(2)} The Fourier transform of the total correlation function is calculated. Invoking the spherical symmetry, $H(k)=(4\pi/k)\int_0^{\infty}rh(r)\sin{(kr)}dr$ emerges leading to $H(k_i)=(4\pi\Delta{r}/k_i)\sum_{j=1}^{W}r_jh(r_j)\sin{[(\pi/W)(j-\frac{1}{2})i]}$ for the discrete sine transform of a space resolution equal to the bin width (to avoid interpolations). In the above, $W$ is the histogram bin number, $\Delta{k}=\pi/(N\Delta{r})$, $k_i=\imath\Delta{k}$, $r_j=j\Delta{r}-\Delta{r}/2$, $r=\{r_i\}$ and $k=\{k_i\}$. Fast Fourier Transform (FFT) algorithms are employed in order to reduce the computational cost. \textbf{(3)} The Fourier transform of the direct correlation function is computed. By Fourier transforming the OZ equation and solving for $C(k)$, we obtain $C(k)=H(k)/[1+nH(k)]$. \textbf{(4)} The direct correlation function is calculated from the inverse Fourier transform. Invoking spherical symmetry, we acquire the expression $c(r)=[1/(2\pi^2r)]\int_0^{\infty}kC(k)\sin{(kr)}dk$ that ultimately leads to the discrete inverse sine transform $c(r_i)=[\Delta{k}/(2\pi^2r_i)]\{\sum_{j=1}^{W-1}k_jC(k_j)\sin{[(\pi/W)j(i-\frac{1}{2})]}+\mathrm{R}_{W,i}\}$ with $\mathrm{R}_{W,i}=[(-1)^{i-1}/2]\,\,k_WC(k_W)$ for the residue. FFT algorithms are again employed. \textbf{(5)} The bridge function is computed. The closure equation is solved for the bridge function leading to $B(r)=\ln[g(r)]-h(r)+c(r)+\beta{u}(r)$.

Inside the correlation void, that can be loosely defined as $\displaystyle\mathrm{arg}_{r}\{g(r)\ll1\}$ or $\displaystyle\mathrm{arg}_{r}\{g(r)\simeq0\}$, particle encounters are extremely rare but the probability remains finite. As a consequence, the histogram method could, for instance, lead to either $g(r)=10^{-8}$ or to $g(r)=10^{-12}$ due to the poor statistical sampling. The mathematical structure of the OZ integral equation ensures that this enormous statistical uncertainty does not propagate up to the direct correlation function. However, because of the logarithm that is present in the OZ closure equation which becomes $B(r)=\ln[g(r)]+1+c(r)+\beta{u}(r)$ at very short distances, it strongly impacts the bridge function leading to a significant $-8$ or $-12$ contribution for this example. In conclusion, the unavoidable insufficient statistics within the correlation void suggest that the OZ inversion method is only effective for the computation of the bridge function at intermediate and long ranges. For a given interaction potential, the validity limit mainly depends on the state point of interest, the overall statistics (number of particles and number of uncorrelated configurations) and the desired accuracy.

\subsection{The production runs}\label{subProductionRuns}

\noindent The production runs for the extraction of the radial distribution functions, as well as the tracing of the isomorphic curves, were carried out on graphics cards with the RUMD open-source software\,\cite{RUMDref1}. A small number of test runs were performed with the LAMMPS package\,\cite{RUMDref2}.

In the production runs dedicated to $g(r)$ extraction; the (canonical) NVT MD simulations utilize the shifted-force cut-off method with the Yukawa pair potential truncated at $r_{\mathrm{cut}}=10d$ and the time-step employed for the propagation of equations of motion is $\Delta{t}/\tau=2.5\times10^{-3}$. The MD equilibration time is $2^{20}\Delta{t}$, the statistics duration is $2^{23}\Delta{t}$ and the configuration saving period is $2^{7}\Delta{t}$ leading to $M=2^{16}=65536$ for the statistically independent configurations. The particle number is $N=54872$ leading to $60d$ for the cubic simulation box length. The bin width size of the histogram method is $\Delta{r}/d=0.002d$.

The configuration saving period was selected so that uncorrelated radial distribution functions are always extracted, see section \ref{subErrorStatisticalRadial}. The combination of the statistics duration and the number of particles was selected so that sufficient pair correlation sample sizes are collected even up to $r=1.25d$, see section \ref{subErrorStatistical}. The size of the histogram bin width was selected so that grid errors are much smaller than statistical errors, see section \ref{subErrorGrid}. A large number of test runs were carried out in order to choose a near-optimal cut-off method, truncation radius and MD time-step. These runs were carried out at the YOCP state point $\Gamma_{\mathrm{ISO}}^{\mathrm{OCP}}=160,\,\kappa=1.0$ for which the bridge function exhibits the highest sensitivity to uncertainties in the radial distribution function, see section \ref{SyntheticUncertainties}. Finally, some test runs were performed with the RUMD and LAMMPS softwares for the same YOCP state points and for identical simulation settings as a validation check.

\subsection{The numerical results}\label{subNumericalResults}

\begin{figure*}
	\centering
	\includegraphics[width=7.05in]{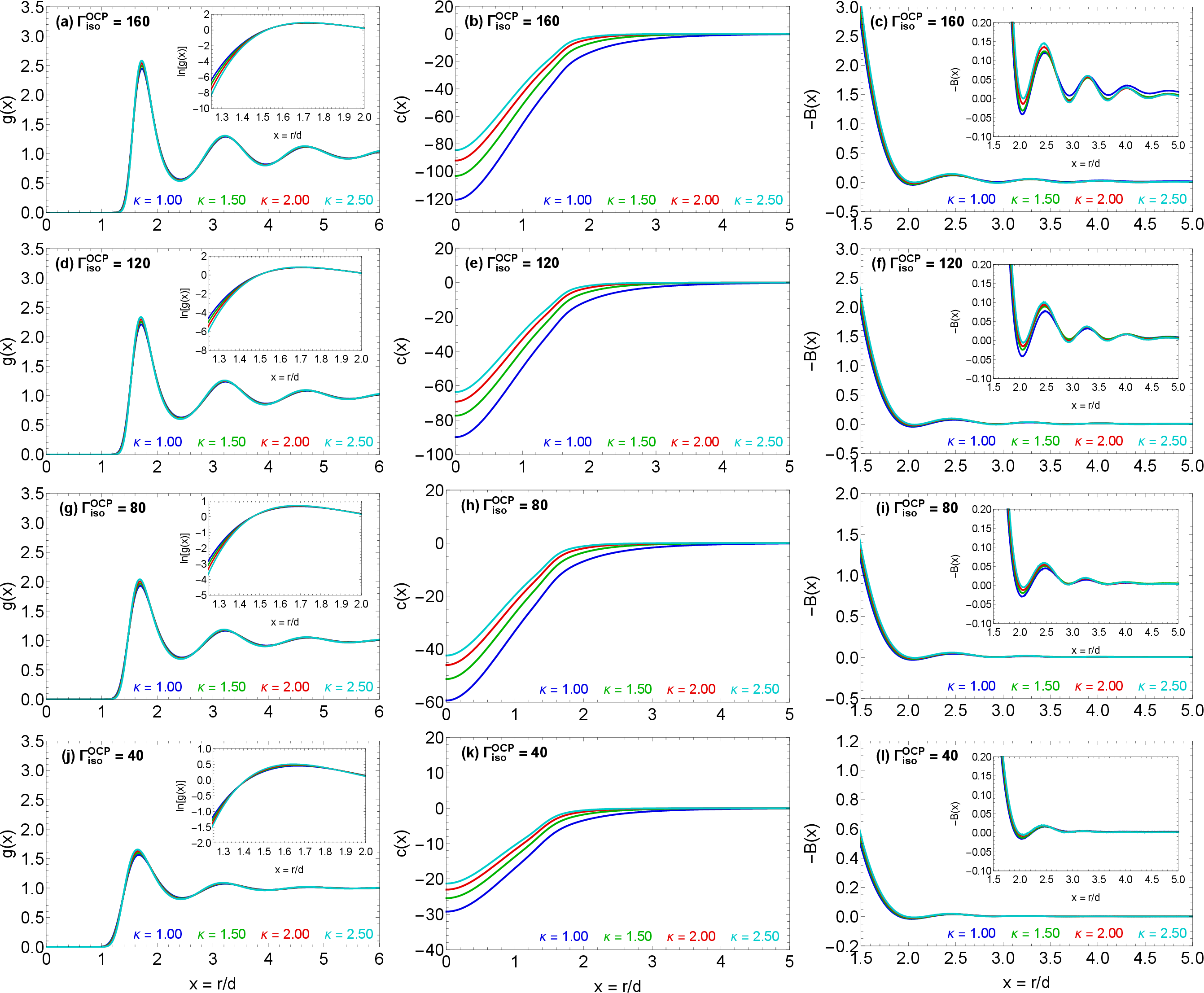}
	\caption{Static correlation functions resulting from application of the Ornstein-Zernike inversion method for the YOCP with radial distribution function input from NVT MD simulations (see the production runs). Results for all the $16$ state points of interest. (a,\,d,\,g,\,j) The radial distribution function and the potential of mean force for the $4$ members of the $\Gamma_{\mathrm{ISO}}^{\mathrm{OCP}}=160,\,120,\,80,\,40$ isomorphs, respectively. (b,\,e,\,h,\,k) The direct correlation function for the $4$ members of the $\Gamma_{\mathrm{ISO}}^{\mathrm{OCP}}=160,\,120,\,80,\,40$ isomorphs, respectively. (c,\,f,\,i,\,l) The bridge function for the $4$ members of the $\Gamma_{\mathrm{ISO}}^{\mathrm{OCP}}=160,\,120,\,80,\,40$ isomorphs, respectively.}\label{fig:invariance_160_120_80_40}
\end{figure*}

\noindent The sequential output of the OZ inversion method, \emph{i.e.} $g(r)\to{c}(r)\to{B}(r)$, is illustrated in figure \ref{fig:invariance_160_120_80_40} for the $4$ isomorphic curves and $16$ YOCP state points of interest.

The radial distribution functions $g(r/d)$ along each isomorph are illustrated in the first panel within the interval $r/d\leq6$, see subfigures \ref{fig:invariance_160_120_80_40}\,(a),\,(d),\,(g),\,(j). In addition, the potentials of mean force $-\ln{[g(r)]}$ are plotted within the range $1.25\leq{r}/d\leq2$ in the respective insets. As demonstrated from earlier investigations of the YOCP\,\cite{isoYOCPg}, the radial distribution function is a strongly invariant quantity along any isomorphic curve with the exception of a narrow interval around the first maximum. This invariance holds to a very good approximation outside the correlation void but is rapidly distorted at short distances. The local variance becomes especially apparent when inspecting the potentials of mean force for $r/d\lesssim1.5$. To be more concrete, the $g(r)$ deviations between isomorphic state points reach two orders of magnitude at $r=1.25d$ for $\Gamma_{\mathrm{ISO}}^{\mathrm{OCP}}=160$ and this trend is expected to get further augmented at shorter distances. This behavior does not contradict the basic property of R-simple systems which states that they possess approximate invariant structural properies in reduced $r/d$ units, because it manifests itself in short distances where the radial distribution function can be approximated with zero and its exact values are inconsequential. In other words, this behavior concerns ultra-rare structural configurations that are physically insignificant. Actually, the observed correlation void variance is a direct consequence of the asymptotic limit of a theorem derived by Widom, which states that $g(r)$ becomes proportional to $\exp{[-\beta{u}(r)]}$ as $r\to0$\,\cite{Widomthe}.

The direct correlation functions $c(r/d)$ along each isomorph are illustrated in the second panel for the interval $r/d\leq5$, see the subfigures \ref{fig:invariance_160_120_80_40}(b),(e),(h),(k). It is evident that the direct correlation function is a strongly variant quantity everywhere. This could be expected from the reduced excess inverse isothermal compressibility relation $\bar{\mu}_{\mathrm{T}}=-n\int{c}(r)d^3r$ and the fact that $\bar{\mu}_{\mathrm{T}}$ is variant as a thermodynamic quantity that involves second order volume derivatives\,\cite{isomdic3} as well as from the exact asymptotic limit $c(r\to\infty)=-\beta{u}(r)$\,\cite{Hansenbo}. It is worth noting that direct correlation functions reach their asymptotic limit much faster than other static correlation functions. This takes place prior to $r/d=2$, close to the foot of the $c(r)$ curve where the slope exhibits a rapid change. This observation justifies the satisfactory performance of the soft mean spherical approximation for the YOCP\,\cite{SMSAmina,SMSAminb}.

The bridge functions $B(r/d)$ along each isomorph are illustrated in the third panel for the interval $1.5\leq{r}/d\leq5$, see subfigures \ref{fig:invariance_160_120_80_40}\,(c),\,(f),\,(i),\,(l). The insets feature a magnification of the oscillatory bridge function pattern. It is evident that the bridge function is a strongly invariant quantity along any isomorph curve in the whole range across which it can be accurately computed with the OZ inversion method. In contrast to the radial distribution function for which there are strong variant features within the correlation void and for which longer range weak variant features are concentrated in the first maximum vicinity, the bridge function variant features appear to be uniformly spread in the whole computation range. In the long range, the observed invariance is justified by the asymptotic behavior of the bridge function. Substituting for $c(r)\simeq-\beta{u}(r)$ and $g(r)=h(r)+1$ in the closure equation, we have $B(r)=\ln{[h(r)+1]}-h(r)$. Taylor expanding the logarithm with respect to $h(r)\simeq0$ and retaining up to the second order term, we end up with $B(r)=-(1/2)h^2(r)$. In the intermediate range, the observed invariance is in accordance with the h-bond expansion that formally defines the bridge function through the infinite series $B(r)=\sum_{i=2}^{\infty}b_i(r)n^{i}$ where the unknown weighting functions $b_i(r)$ are given by multiple integrals only involving the total correlation function\,\cite{hbondbok}. Since the integration space for the weighting function of the $n^{i}$ term is $\mathds{R}^{3i}$, the introduction of reduced units $r/d$ or $r/\Delta$ means that all the powers of the density vanish for each term, thus leading to $B(r/d)=\sum_{i=2}^{\infty}b_i(r/d)$.

To sum up, for all YOCP isomorphic lines, a high level of bridge function invariance has been observed. However, the small deviations between the bridge functions of YOCP state points that belong to the same isomorph might be comparable to the omnipresent uncertainties in bridge function determination. Therefore, in order to accurately quantify the level of isomorph invariance of the bridge function in the intermediate and long ranges, a detailed analysis of all different uncertainty sources is required. This is pursued in sections \ref{SyntheticUncertainties} and \ref{RealUncertainties}.

\section{Sensitivity studies}\label{SyntheticUncertainties}

\noindent Bridge function uncertainties originate from uncertainties in the simulation-extracted radial distribution functions that propagate through all the steps of the OZ inversion method. For distances far from the edge of the correlation void that can be loosely defined by $g(r)\ll1$, the induced direct correlation function uncertainties are dominant. On the other hand, in the vicinity of the correlation void, the radial distribution function uncertainties themselves are the most crucial. Only the former regime is of relevance for the present work, since the OZ inversion method fails within the correlation void.

In modern computer simulations, where large numbers of particles and long observation times are routinely feasible, errors in extracted radial distribution functions are negligible. However, as discussed earlier, it is well-known that bridge functions are highly sensitive to radial distribution function uncertainties. Rigorous quantification of the sensitivity degree is complicated due to the non-linear non-local nature of the exact functional that connects the two quantities. Useful insights can be gained by inserting small controlled perturbations in the extracted $g(r)$ and then documenting their effect in the computed $B(r)$. The artificial radial distribution function has the general form $g(r)=g_{\mathrm{MD}}(r)\times[1+\Delta{g}(r)]$. Multiplicative perturbations are preferred over additive perturbations of the form $g(r)=g_{\mathrm{MD}}(r)+\Delta{g}(r)$, because they always slightly distort the correlation void by construction and because they emerge in the treatment of finite size errors.

\begin{figure*}
	\centering
	\includegraphics[width=6.4in]{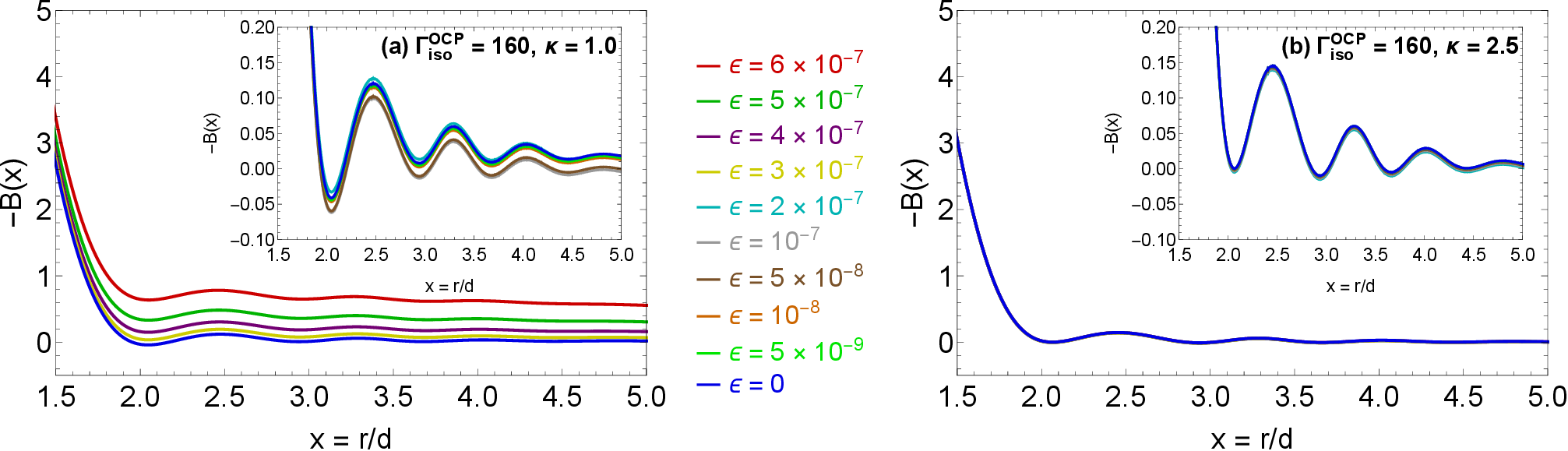}
	\caption{Sensitivity of the YOCP bridge function to \emph{artificial aperiodic uncertainties} in the radial distribution function at two state points along the isomorph $\Gamma_{\mathrm{ISO}}^{\mathrm{OCP}}=160$. (a) For $\kappa=1.0$; error magnitudes $\epsilon>2\times10^{-7}$ lead to large perturbations in the bridge function, error magnitudes $\epsilon>10^{-8}$ lead to small but observable perturbations and error magnitudes $\epsilon<5\times10^{-9}$ are required for the bridge function to become insensitive. (b) For $\kappa=2.5$; error magnitudes $\epsilon<6\times10^{-7}$ suffice for the bridge function to become insensitive to perturbations. There is a dramatic dependence of the sensitivity on the screening parameter.}\label{fig:sensitivity_aperiodic}
\end{figure*}

\begin{figure*}
	\centering
	\includegraphics[width=6.4in]{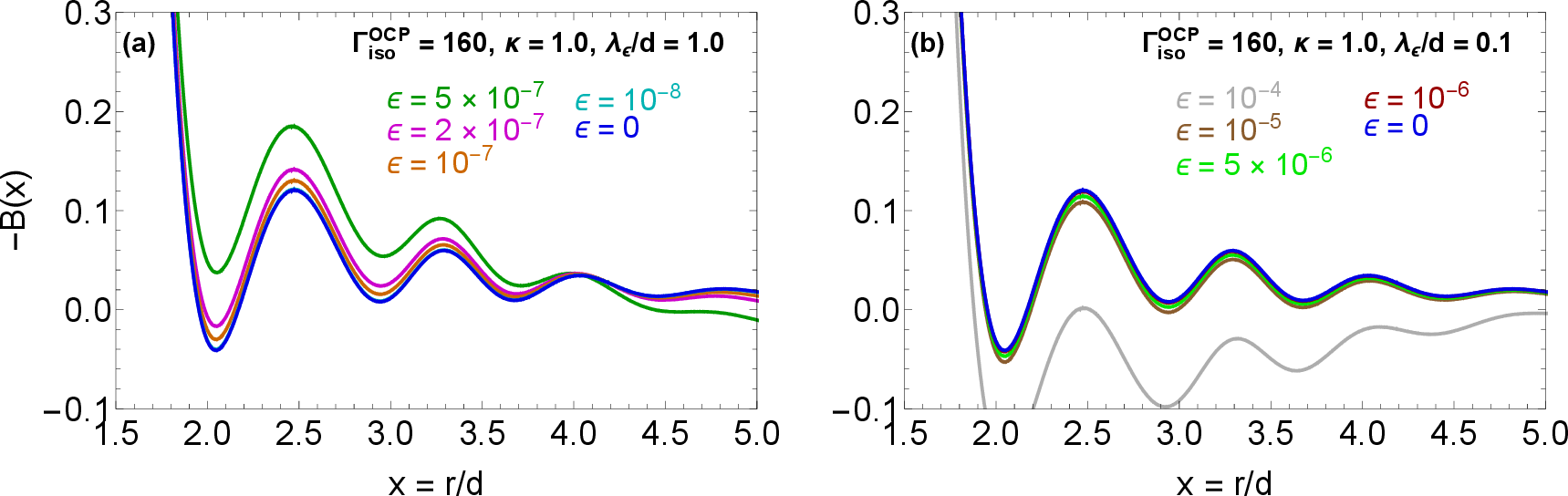}
	\caption{Sensitivity of the YOCP bridge function to \emph{artificial periodic uncertainties} in the radial distribution function at the state point $\Gamma_{\mathrm{ISO}}^{\mathrm{OCP}}=160,\,\kappa=1.0$ for different wavelengths. (a) For $\lambda_{\mathrm{\epsilon}}=d$, error magnitudes $\epsilon<10^{-7}$ lead to very small bridge function perturbations. (b) For $\lambda_{\mathrm{\epsilon}}=0.1d$, error magnitudes $\epsilon<10^{-5}$ lead to very small bridge function perturbations. There is a strong dependence of the sensitivity on the error wavelength.}\label{fig:sensitivity_periodic}
\end{figure*}

The propagation of two types of artificial errors was investigated for all the $16$ YOCP state points of interest. \emph{Aperiodic uncertainties} of the form $g(r)=g_{\mathrm{MD}}(r)\times[1+\epsilon]$ with the following values of the error amplitude probed: $\epsilon=(6,\,5,\,4,\,3,\,2,\,1,\,0.5,\,0.1,\,0.05,\,0.01)\times10^{-7}$. \emph{Periodic uncertainties} of the form $g(r)=g_{\mathrm{MD}}(r)\times[1+\epsilon\cos{(r/\lambda_{\mathrm{\epsilon}})}]$ with the wavelengths $\lambda_{\mathrm{\epsilon}}/d=(1,\,0.1,\,0.01,\,0.001)$ probed and five targeted error amplitudes for each wavelength: $\epsilon=10^{-8}-10^{-6}$ for $\lambda_{\mathrm{\epsilon}}=d$, $\epsilon=10^{-6}-10^{-3}$ for $\lambda_{\mathrm{\epsilon}}=0.1d$, $\epsilon=10^{-5}-10^{-2}$ for $\lambda_{\mathrm{\epsilon}}=0.01d$ and $\epsilon=5\times10^{-4}-10^{-2}$ for $\lambda_{\mathrm{\epsilon}}=0.001d$. The bridge functions originating from the unperturbed and perturbed MD-extracted radial distribution functions were compared in order to deduce the bridge function sensitivity. Some characteristic examples are illustrated in figures \ref{fig:sensitivity_aperiodic} and \ref{fig:sensitivity_periodic}.

From this numerical investigation, the following conclusions can be drawn: \textbf{(a)} Regardless of the state point and for constant error magnitude $\epsilon$, the bridge function is mainly sensitive to aperiodic uncertainties in the radial distribution function. \textbf{(b)} Regardless of state point and for constant error magnitude $\epsilon$, as the wavelength $\lambda_{\mathrm{\epsilon}}$ of periodic uncertainties in the radial distribution function decreases, the sensitivity of the bridge function becomes progressively weaker. \textbf{(c)} As $\kappa$ increases for constant normalized coupling parameters $\Gamma/\Gamma_{\mathrm{m}}$, the sensitivity of the bridge function to periodic and aperiodic uncertainties dramatically decreases. Thus, bridge function uncertainties are larger near the OCP limit. \textbf{(d)} As $\Gamma/\Gamma_{\mathrm{m}}$ increases for a constant screening parameter $\kappa$, the sensitivity of the bridge function to periodic and aperiodic uncertainties increases. Hence, bridge function uncertainties are larger near the melting line. \textbf{(e)} The study of aperiodic errors for the $(\Gamma_{\mathrm{ISO}}^{\mathrm{OCP}}=160,\,\kappa=1.0)$ YOCP state point corresponds to the worst case scenario, where $\epsilon<5\times10^{-9}$ was required for bridge functions to be indistinguishable.

The artificial error studies demonstrate the well-known fact that very accurate simulation-extracted radial distribution functions should be available for reliable computation of the bridge function. It is evident though that the desired level of accuracy depends strongly on the YOCP state point. The aforementioned trends will be invoked in the discussion of statistical errors and their variations with the state point, in the analysis of grid errors and the existence of a near-optimal bin width, in the investigation of explicit finite size errors and the necessity for their correction, as well as in the rationale behind the discarding of implicit finite size errors.

\section{Propagation of uncertainties}\label{RealUncertainties}

\noindent Five types of uncertainties are relevant to investigations of the isomorph invariance of bridge functions:
\begin{enumerate}[leftmargin=+4mm,topsep=1.0ex,itemsep=-0ex]
    \item \emph{statistical errors} due to the finite simulation duration,
    \item \emph{grid errors} due to the finite histogram bin width,
    \item \emph{size errors} due to the finite simulated particle number,
    \item \emph{tail errors} due to the finite simulation box length,
    \item \emph{isomorphic errors} due to excess entropy mismatches.
\end{enumerate}
The first four uncertainties refer to the propagation of radial distribution function uncertainties and are relevant for all bridge function studies. The last uncertainty refers to the propagation of isomorphic state point uncertainties and is only relevant for bridge function studies which test the degree of invariance along isomorphic curves.

In the rich literature of bridge function extraction by computer simulations, finite size errors and tail errors are usually discussed, since rigorous procedures for their correction have been developed. On the other hand, statistical errors and grid errors are rarely analyzed in depth. Notable exceptions to this norm are the detailed uncertainty analysis performed by Poll and collaborators for the OCP bridge functions\,\cite{PollOCPB} as well as by Kolafa and collaborators for the hard sphere bridge functions\,\cite{bridgHS1}. In the present investigation, a detailed quantification of uncertainties is crucial in order to correctly attribute the physical origin of the small deviations which are observed between the bridge functions of state points that belong to the same YOCP isomorph. To our knowledge, the following uncertainty study is the most meticulous analysis to be reported in the literature thus far.

\subsection{Statistical errors}\label{subErrorStatistical}

\noindent Statistical errors emerge in the extraction of any thermodynamic, structural or dynamic property from computer simulations as a consequence of their finite duration\,\cite{statgen1}. For ergodic systems, the thermodynamic ensemble average of physical quantities can be substituted by their long time average. However, the latter can only be practically approximated in a truncated series form, \emph{i.e.}
\begin{align*}
\int\,a(\boldsymbol{r}^{N},\boldsymbol{p}^{N})f_{N}(\boldsymbol{r}^{N},\boldsymbol{p}^{N})d^3r^{N}d^3p^N&=\lim_{\tau\to\infty}\frac{1}{\tau}\int_0^{\tau}a(t)dt\\&\simeq\frac{1}{K}\sum_{i=1}^{K}a(i\Delta{t})
\end{align*}
with $f_{N}(\boldsymbol{r}^{N},\boldsymbol{p}^{N})$ the $N-$particle distribution function, $a$ an arbitrary physical quantity, $\Delta{t}$ the simulation time step, $K$ the total number of time steps in the equilibrium state. Due to the inherent fluctuations of $a(i\Delta{t})$ at each time step and the finite value of $K$, statistical uncertainties arise in the average value of $a$. In the following, we shall denote thermodynamic ensemble averages with $\langle...\rangle$ and simulation averages with $\langle...\rangle_{K}$ where $K$ is the number of samples.

\subsubsection{Level of radial distribution functions}\label{subErrorStatisticalRadial}

\noindent In order to calculate the statistical uncertainties in the radial distribution function, the number of time steps that is required for radial distribution functions to become uncorrelated at all distances should be first identified. This was accomplished with the application of the block averaging method proposed by Flyvbjerg and Petersen\,\cite{statgen2} at all distances and led to the conclusion that $64$ MD time steps are required, regardless of the YOCP state point. Adding a safety margin, after the equilibration phase, the configurations were saved every $128$ time steps for the extraction of $g(r)$ in the production runs. It is evident that any larger period of configuration saving would unnecessarily increase the computational cost. Given the lack of time-correlations, well-known statistical formulas can be employed for the average radial distribution function and its standard deviation,
\begin{align*}
&g(r)=\langle{g}(r)\rangle_{M}=\frac{1}{M}\sum_{i=1}^{M}g(r,i\Delta{t})\,,\\
&\sigma[g(r)]=\sqrt{\frac{1}{M(M-1)}\sum_{i=1}^{M}\left[g(r,i\Delta{t})-\langle{g}(r)\rangle_{M}\right]^2}\,,
\end{align*}
with $M$ the total number of saved (uncorrelated) configurations ($M=65536$) and where $\sigma$ denotes the standard deviation of the average or standard error of the mean.

\begin{figure}
	\centering
	\includegraphics[width=3.35in]{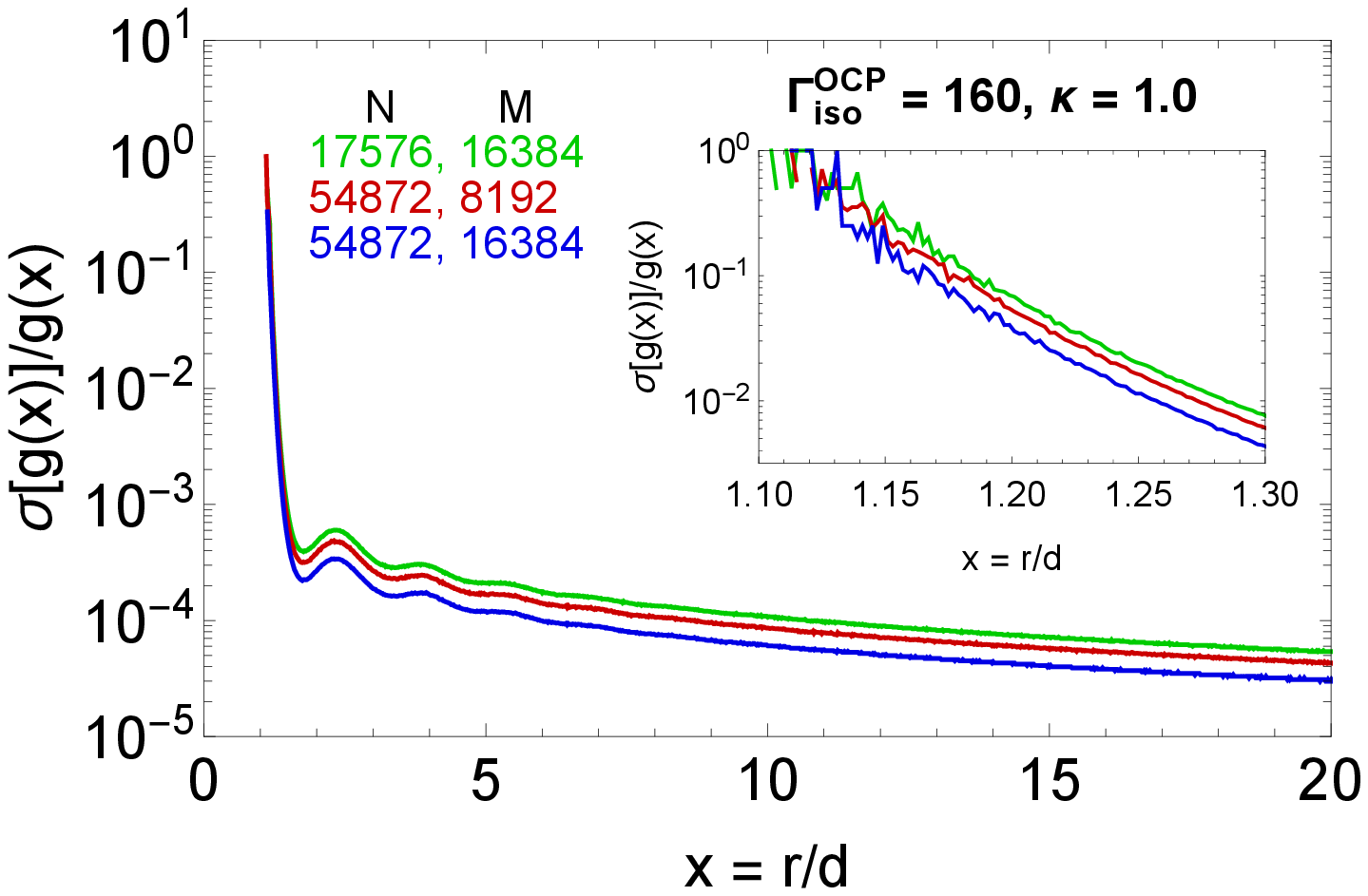}
	\caption{\emph{Relative standard errors} in the MD-extraction of radial distribution function at the YOCP state point $\Gamma_{\mathrm{ISO}}^{\mathrm{OCP}}=160,\,\kappa=1.0$. Results for varying particle number ($N$) and uncorrelated configuration number ($M$). Here,
     $r_{\mathrm{cut}}=16d$ and $T_{\mathrm{save}}=64\Delta{t}$ was employed for the potential cut-off and configuration saving period instead of the standard $r_{\mathrm{cut}}=10d$, $T_{\mathrm{save}}=128\Delta{t}$ of the production runs. See the inset for the insufficient statistics within the correlation void.}\label{fig:statistical_error_rdf}
\end{figure}

The uncertainty $\sigma[g(r)]$ depends on the simulation duration through the sample size $M$ and the particle number through the inherent $g(r,i\Delta{t})$ fluctuation level, since for a single configuration the total $g(r,i\Delta{t})$ statistics are equal to the $N(N-1)$ number of pairs. Figure \ref{fig:statistical_error_rdf} illustrates these dependencies for the relative standard error $\sigma[g(r)]/g(r)$. Notice that the relative standard error exhibits a divergent behavior at short distances, since close particle encounters are extremely rare. In particular, distances within the correlation void are so poorly sampled that the error in $g(r)$ becomes comparable to the $g(r)$ average value prior to one inter-particle distance $d$. Our production runs were designed in a manner that guarantees $\sigma[g(r)]/g(r)<0.01$ up to $r=1.25d$.

\subsubsection{Level of bridge functions}\label{subErrorStatisticalBridge}

\noindent In principle, knowledge of $\sigma[g(r)]$ should allow for a determination of $\sigma[B(r)]$ by standard error propagation analysis. However, the non-locality of the OZ equation prohibits such calculations, since $\sigma[c(r)]$ depends on the $g(r)$ values at all possible distances that are naturally correlated with each other. Nevertheless, the application of error analysis to the closure equation alone can be used in order to reaffirm the inapplicability of OZ inversion at short distances. Substituting for $g(r)\to{g}(r)+\sigma[g(r)]$, $B(r)\to{B}(r)+\sigma[B(r)]$, $c(r)\to{c}(r)+\sigma[c(r)]$, linearizing with respect to the small uncertainties and disposing the averages via the unperturbed closure equation, we obtain
\begin{equation*}
\sigma[B(r)]=\frac{\sigma[g(r)]}{g(r)}-\sigma[g(r)]+\sigma[c(r)]\,.
\end{equation*}
The first term represents the relative standard error in the radial distribution function and stems from the logarithmic term of the closure equation. As illustrated in figure \ref{fig:statistical_error_rdf}, this term rapidly diverges at short distances where it should overcome contributions from other terms, thus demonstrating that $B(r)$ cannot be reliably computed inside the correlation void with the OZ inversion method.

\begin{figure*}
	\centering
	\includegraphics[width=6.4in]{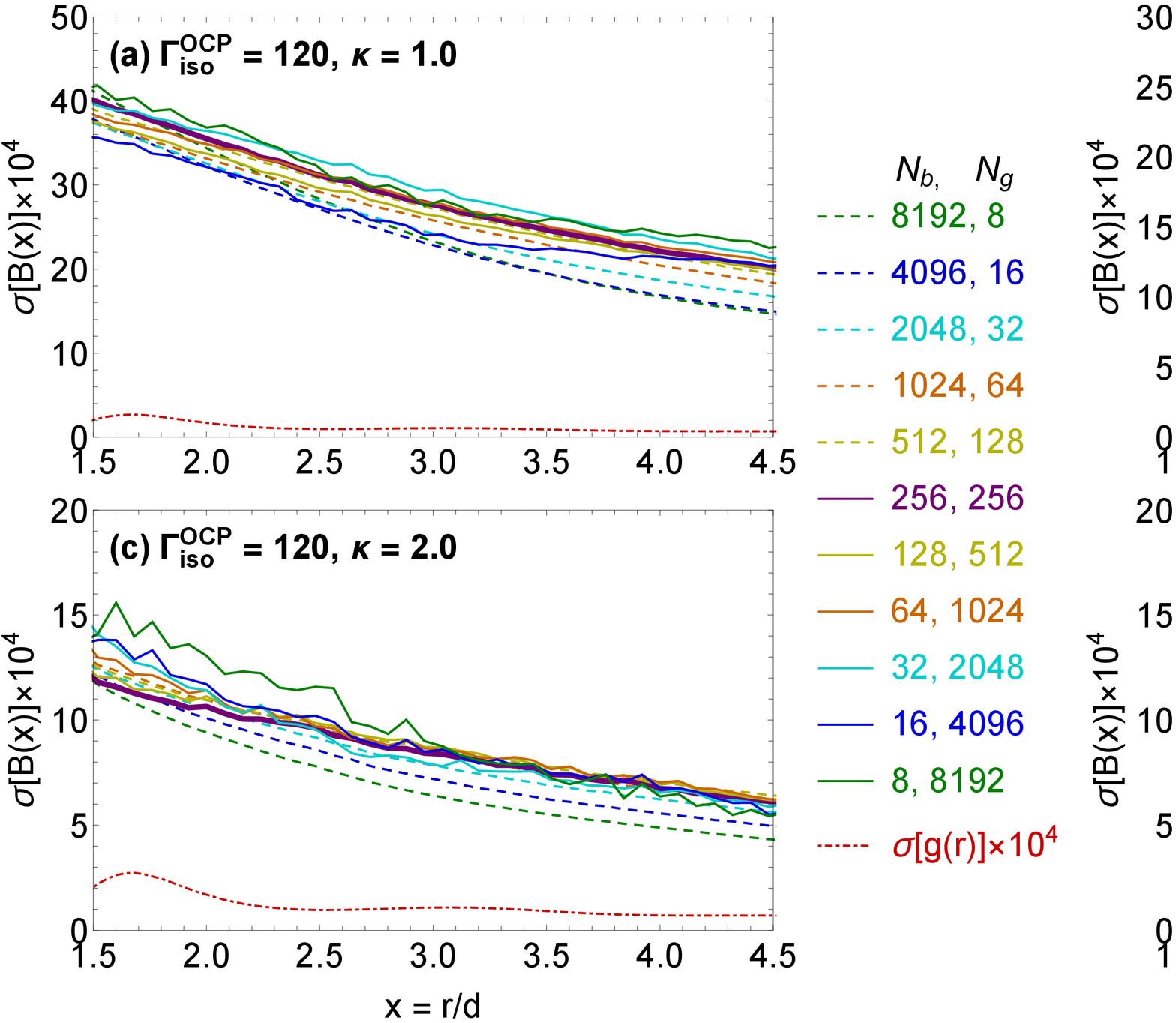}
	\caption{\emph{Statistical errors} in the computation of the YOCP bridge function along the isomorph $\Gamma_{\mathrm{ISO}}^{\mathrm{OCP}}=120$ ($\kappa=1.0,1.5,2.0,2.5$). Determination for $11$ different symmetrical combinations of the block number and sub-block configuration number;
     $(N_{\mathrm{b}},\,N_{\mathrm{g}})=\{(8192,8),(4096,16),(2048,32),(1024,64),(512,128),(256,256),(128,512),(64,1024),(32,2048),(16,4096),(8,8192)\}$. Statistical errors from the MD-extraction of the respective radial distribution functions are also provided. The strong overlapping of the statistical errors for the combinations $(N_{\mathrm{b}},\,N_{\mathrm{g}})=\{(512,128),(256,256),(128,512)\}$ justifies the choice of $(N_{\mathrm{b}},\,N_{\mathrm{g}})=(256,256)$. Note the monotonic decrease of the statistical errors as the screening parameter increases.}\label{fig:statistical_error}
\end{figure*}

In view of the calculation of the $g(r)$ statistical errors, it would seem appropriate to compute the bridge function $B(r,i\Delta{t})$ for all extracted $g(r,i\Delta{t})$ and then to employ
\begin{align*}
&B(r)=\langle{B}(r)\rangle_{M}=\frac{1}{M}\sum_{i=1}^{M}B(r,i\Delta{t})\,,\\
&\sigma[B(r)]=\sqrt{\frac{1}{M(M-1)}\sum_{i=1}^{M}\left[B(r,i\Delta{t})-\langle{B}(r)\rangle_{M}\right]^2}\,.
\end{align*}
These expressions would lead to erroneous results both for the average bridge function and for its standard deviation. The reason is that the bridge function and direct correlation function lack microscopic representation, in contrast to the radial distribution function and structure factor. These static correlation functions are only defined in the thermodynamic limit and their reliable computation from OZ inversion requires a sufficiently smooth radial distribution function as input. Thus, strictly speaking, the quantity $B(r,i\Delta{t})$ does not make physical sense and should not be employed in the calculation of the average bridge function and its standard deviation. In fact, since the bridge function is a highly non-linear functional of the radial distribution function, there are large deviations between $\langle{B}[g(r,i\Delta{t})]\rangle_{M}$ and $B[\langle{g}(r,i\Delta{t})\rangle_{M}]$ with the second relation representing the correct way to calculate the average bridge function.

Based on these observations, the following procedure was devised to estimate the propagation of statistical uncertainties to bridge functions. \textbf{\emph{First}}, the total dataset of M uncorrelated configurations was divided into $N_{\mathrm{b}}$ blocks each containing $N_{\mathrm{g}}$ configurations ($M=N_{\mathrm{b}}\times{N}_{\mathrm{g}}$). The $N_{\mathrm{g}}$ blocking ensured the extraction of sufficiently smooth radial distribution functions and the $N_{\mathrm{b}}$ blocking ensured large sample sizes for the calculation of the statistical deviations. \textbf{\emph{Then}}, for each block, the block-averaged radial distribution function and bridge function were computed: $\langle{g}_{i}(r)\rangle_{N_{\mathrm{g}}}=(1/N_{\mathrm{g}})\sum_{j=1}^{N_{\mathrm{g}}}g(r,j\Delta{t})$,\,$B_{i}(r)=B[\langle{g}_{i}(r)\rangle_{N_{\mathrm{g}}}]$. It is worth pointing out that average bridge function resulting from $(1/N_{\mathrm{b}})\sum_{i=1}^{N_{\mathrm{b}}}B_i(r)$ is different from the true average bridge function $B[\langle{g}(r)\rangle_{M}]$ and also depends on the values chosen for $N_{\mathrm{b}},N_{\mathrm{g}}$. \textbf{\emph{Afterwards}}, the standard deviation of the mean was computed, at each point, by
\begin{equation*}
\sigma[B(r)]=\sqrt{\frac{1}{N_{\mathrm{b}}(N_{\mathrm{b}}-1)}\sum_{i=1}^{N_{\mathrm{b}}}\left\{B_{i}(r)-B[\langle{g}(r)\rangle_{M}]\right\}^2}\,.
\end{equation*}
\textbf{Finally}, the $(N_{\mathrm{b}},\,N_{\mathrm{g}})$ values should be determined. The $N_{\mathrm{g}}$ value should be as large as possible for $\langle{g}_{i}(r)\rangle_{N_{\mathrm{g}}}$ to be sufficiently smooth, otherwise $B_{i}(r)$ would be unphysical. The $N_{\mathrm{b}}$ value should be as large as possible for the $B_{i}(r)$ sample size to be sufficiently large, otherwise the estimate of $\sigma[B(r)]$ would be unreliable. Within the constraint of $M=N_{\mathrm{b}}\times{N}_{\mathrm{g}}$ and assuming that the above requirements are of equal significance, then the near-optimal values are $N_{\mathrm{g}}=N_{\mathrm{b}}=\sqrt{M}$ or $N_{\mathrm{g}}=N_{\mathrm{b}}=256$.

The above choice for the $N_{\mathrm{b}},N_{\mathrm{g}}$ combination was also confirmed to be near-optimal by the objective empirical analysis described below. For all $16$ YOCP state points of interest, the statistical uncertainties in the bridge function were determined for $11$ symmetric $(N_{\mathrm{b}},N_{\mathrm{g}})$ combinations. The statistical error was observed to strongly fluctuate for combinations that featured low values of $N_{\mathrm{b}},N_{\mathrm{g}}\lesssim64$, but remained nearly constant for the combinations $(N_{\mathrm{b}},\,N_{\mathrm{g}})=\{(512,128),(256,256),(128,512)\}$, see figure \ref{fig:statistical_error} for an example. Since $(N_{\mathrm{b}},\,N_{\mathrm{g}})=(256,256)$ lies at the centre of the stability neighbourhood regardless of the state point, it was selected for the quantification of statistical errors. It should be emphasized that the near-optimal $(N_{\mathrm{b}},\,N_{\mathrm{g}})$ combination not only depends on the number of uncorrelated configurations $M$ but also on the number of simulated particles $N$. Hence, the above choice does not constitute a general recommendation.

Analysis of the statistical errors for all $16$ state points that belong to the $4$ YOCP isomorphs led to the following conclusions, all in accordance with the sensitivity studies. (a) As $\Gamma/\Gamma_{\mathrm{m}}$ increases for a constant screening parameter $\kappa$, the statistical errors in the bridge function increase. (b) As $\kappa$ increases for a constant normalized coupling parameter $\Gamma/\Gamma_{\mathrm{m}}$, the statistical errors in the bridge function decrease. (c) Regardless of the state point, the statistical errors in the bridge function are larger than the statistical errors in the radial distribution function. The largest difference is achieved for $\Gamma_{\mathrm{ISO}}^{\mathrm{OCP}}=160,\,\kappa=1.0$ $(\sim50\times)$ and the smallest difference for $\Gamma_{\mathrm{ISO}}^{\mathrm{OCP}}=40,\,\kappa=2.5$ $(\sim3\times)$.

\subsection{Grid errors}\label{subErrorGrid}

\noindent Grid errors emerge as a consequence of the finiteness of the bin width that is employed in the histogram method extraction of radial distribution functions from computer simulations. In other words, grid errors in the radial distribution function are generated by the fact that only average values in intervals of finite length can be extracted from MD simulations.

In mathematical terms, grid errors originate from the discretization of the microscopic representation of the radial distribution function. It is instructive to revisit the histogram method starting from the $g(r)$ $\delta-$function representation\,\cite{PollOCPB}
\begin{equation*}
ng(r)=\frac{1}{N}\langle\sum_{i=1}^{N}\sum_{j=1}^{N}\delta(\boldsymbol{r}-\boldsymbol{r}_j+\boldsymbol{r}_i)\rangle\,.
\end{equation*}
This expression is integrated within thin spherical shells $(r_{n},r_{n}+\Delta{r})$ with $\Delta{r}\ll{r}_{n}$. The integral of the double series is obtained from the post-processing of the MD trajectories and denoted with $N(r_{n},r_{n}+\Delta{r})$. The integral of $ng(r)$ is evaluated by employing spherical coordinates, expanding the radial distribution function around the effective bin position $x_{n}$ and retaining up to the first order term. Introducing $\Delta{V}=\frac{4}{3}\pi[(r_{n}+\Delta{r})^3-r_{n}^{3}]$, we get
\begin{equation*}
\frac{N(r_{n},r_{n}+\Delta{r})}{Nn\Delta{V}}=g(x_n)+\frac{4\pi}{\Delta{V}}\left[\int_{r_n}^{r_{n}+\Delta{r}}(r-x_n)r^2dr\right]g^{\prime}(x_n)\,.
\end{equation*}
To close the system, we require that the first order term identically vanishes. This results in the conventional histogram method relations
\begin{align*}
&g(x_n)=\frac{N(r_{n},r_{n}+\Delta{r})/N}{\frac{4}{3}\pi{n}[(r_{n}+\Delta{r})^3-r_{n}^{3}]}\,,\\
&x_n=r_n\left\{1+\frac{1}{2}\frac{\Delta{r}}{r_n}+\mathcal{O}\left[\left(\frac{\Delta{r}}{r_n}\right)^2\right]\right\}\simeq{r}_n+\frac{1}{2}\Delta{r}\,,
\end{align*}
that are exact for infinitesimal bin widths $\Delta{r}\to0$ and lead to $g(r)$ errors due to the neglected high-order terms. This generates numerical errors in the calculation of the direct correlation function and thus in the computation of the bridge function. There is clear trade-off between grid errors and statistical errors; smaller bin widths reduce the grid errors but increase statistical errors (certainty in the distances but strong fluctuations in average values), while larger bin widths enhance grid errors but decrease statistical errors (weak fluctuations in the average values due to better grid statistics but uncertainty in distances).

Grid errors should not be confused with numerical errors that stem from the use of discrete Fourier transforms instead of continuous Fourier transforms during the OZ inversion procedure. Within the heuristic assumption of a deterministic radial distribution function (independent of the bin width), Fast Fourier transform routines lead to negligible errors in the YOCP bridge function already for any discretization $\Delta{r}\leq0.1d$. Grid errors stem from the fact that the histogram-extracted $g(r)$ exhibit a non-negligible dependence on bin width variations, \emph{i.e.} it has a stochastic character.

In contrast to statistical errors, grid errors cannot be quantified. Hence, an efficient strategy should be focused on ensuring that sufficiently small bin widths are used so that grid errors are much smaller than statistical errors. The exact value should be decided by an empirical analysis of the dependence of the bridge function on the bin width\,\cite{bridgHS1}. Our investigation revealed that as bin widths become smaller, the bridge functions start to become independent of their value. A near optimal value exists that is close to the largest bin width for which convergence has been achieved, since further reduction would only slightly alter the average bridge function but strongly increase its standard deviation (due to statistical errors).

\begin{figure*}
	\centering
	\includegraphics[width=6.4in]{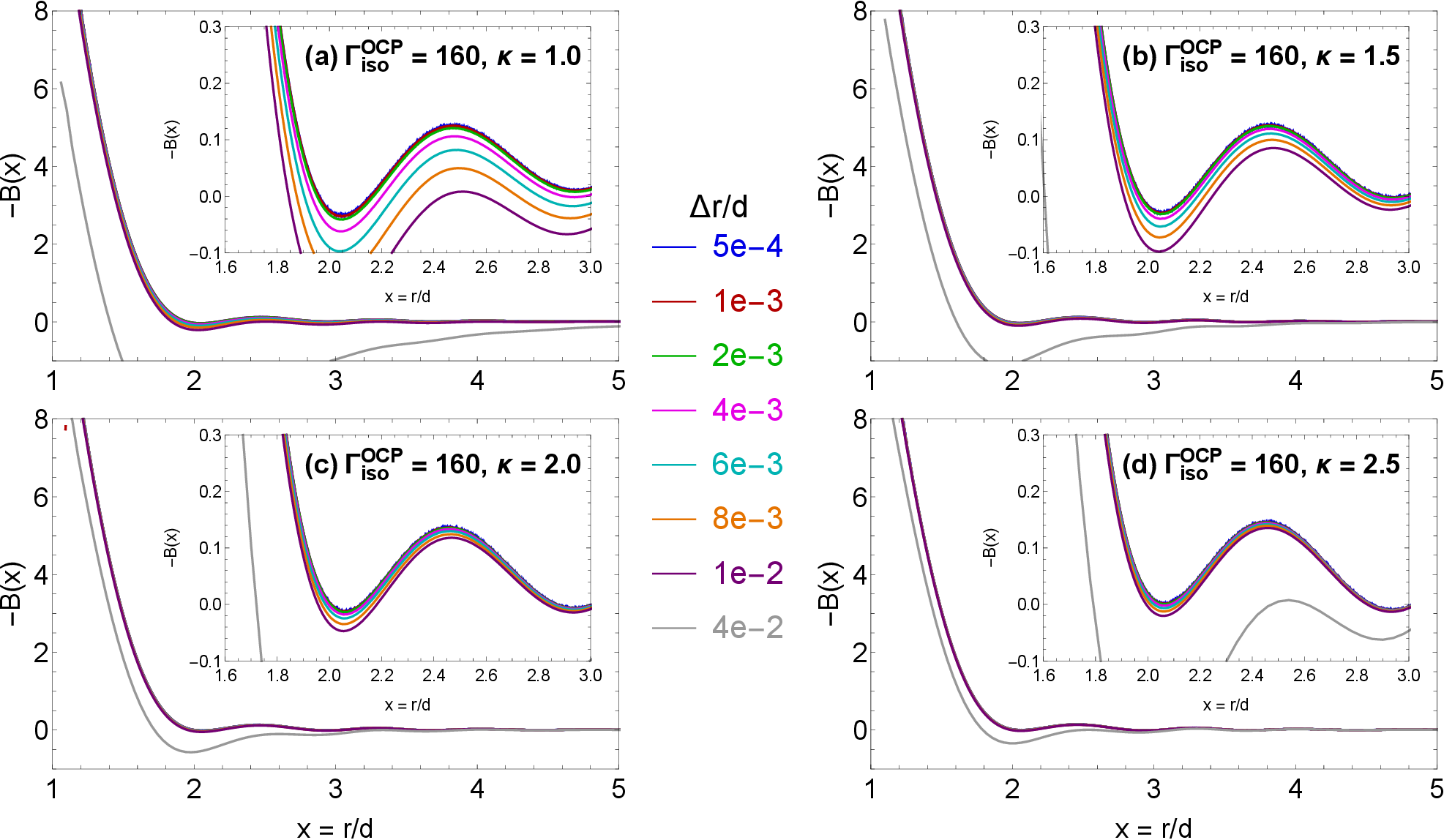}
	\caption{\emph{Grid errors} in the computation of the YOCP bridge function along the isomorph $\Gamma_{\mathrm{ISO}}^{\mathrm{OCP}}=160$ ($\kappa=1.0,1.5,2.0,2.5$). Determination for $8$ different histogram bin widths $\Delta{r}/d=\{0.0005,\,0.001,\,0.002,\,0.004,\,0.006,\,0.008,\,0.01,\,0.04\}$
    It is evident that a near-optimal bin width emerges below which bridge functions overlap. The near-optimal bin width depends slightly on the screening parameter and obtains the rough values $0.002d$ ($\kappa=1.0$), $0.002d$ ($\kappa=1.5$), $0.004d$ ($\kappa=2.0$), $0.006d$ ($\kappa=2.5$). Unless the bin widths are very large ($\Delta{r}/d\gtrsim0.04$), grid errors are rather small and better discerned in the zoomed-in insets.}\label{fig:grid_error}
\end{figure*}

Specifically, for each of the $16$ YOCP state points of interest, the bridge functions were computed from radial distribution functions that were extracted from the same MD simulations but with varying histogram bin widths. Eight bin width values were considered, namely $\Delta{r}/d=\{0.0005,\,0.001,\,0.002,\,0.004,\,0.006,\,0.008,\,0.01,\,0.04\}$. A characteristic example is illustrated in figure\,\ref{fig:grid_error}. Results reveal that: (a) Regardless of the YOCP state point, the bridge functions are always nearly identical for $\Delta{r}\leq0.002d$ bin widths. (b) As $\Gamma/\Gamma_{\mathrm{m}}$ increases with a constant screening parameter $\kappa$, the bridge function depends more strongly on the bin width. (c) As $\kappa$ increases with a constant normalized coupling parameter $\Gamma/\Gamma_{\mathrm{m}}$, the bridge function dependence on the bin width becomes weaker. (d) The near-optimal bin width depends on the YOCP state point. It lies within the interval $\Delta{r}/d=0.002-0.01$ for the probed state points and acquires its smallest value when $\Gamma_{\mathrm{ISO}}^{\mathrm{OCP}}=160,\,\kappa=1.0$. Nevertheless, a state point independent bin width of $\Delta{r}=0.002d$ was preferred. It should be emphasized that the near-optimal bin width should depend strongly on the number of statistics relevant to the extraction of the radial distribution function, \emph{i.e.} on the number of particles and the number of uncorrelated time steps in the MD simulation. Thus, the above choice does not constitute a general recommendation.

The qualitative aspects of the above conclusions were anticipated from the sensitivity study. For instance, let us consider the existence of a near-optimal bin width in greater detail. The grid error can be roughly viewed as an oscillating correction to the radial distribution function with a periodicity of $\lambda_{\mathrm{\epsilon}}\simeq\Delta{r}$. As $\Delta{r}$ gets smaller, then the error $\epsilon$ in the average $g(r)$ becomes larger. As $\lambda_{\mathrm{\epsilon}}$ gets smaller, then the bridge function becomes gradually insensitive to the error $\epsilon$. Consequently, a near-optimal bin width exists where convergence is achieved with these two competing effects cancelling each other and the average bridge function becoming insensitive to the bin width.

\subsection{Finite size errors}\label{subErrorSize}

\noindent Finite size errors ultimately emerge from the finite number of particles considered in computer simulations\,\cite{sizegen1}. In the case of static equilibrium correlations, two types have been identified in the literature\,\cite{sizegen2,sizegen3,sizegen4}; the \emph{explicit (or ensemble) size errors} that emerge in the passage from the canonical ensemble to the grand-canonical ensemble and the \emph{implicit (or anomalous) size errors} that emerge in the passage from a finite simulated system with infinite boundary conditions to an infinite macroscopic system.

\subsubsection{Explicit finite size errors}\label{subErrorSizeExplicit}

\noindent The explicit size effect in NVT MD simulations is a direct consequence of the suppression of particle number fluctuations in the canonical ensemble. The correspondence between the NVT and $\mu$VT radial distribution functions can be derived by expressing the grand canonical two-particle densities via the canonical two-particle densities, Taylor expanding the canonical quantity with respect to the average particle number, retaining up to the second order term and employing the fluctuation relation for the isothermal compressibility. The final expression is the so-called Lebowitz-Percus correction and reads as\,\cite{sizeexp1,sizeexp2,sizeexp3}
\begin{equation*}
g_{\mathrm{c}}(r;n,T)=g_{\mathrm{MD}}(r;n,T)+\frac{\chi_{\mathrm{T}}}{2N}\frac{\partial^2}{\partial{n}^2}\left[n^2g_{\mathrm{MD}}(r;n,T)\right]\,,
\end{equation*}
where $g_{\mathrm{c}}(r;n,T)$ is the corrected radial distribution function, $g_{\mathrm{MD}}(r;n,T)$ the (NVT) MD-extracted radial distribution function, $\chi_{\mathrm{T}}$ the reduced isothermal compressibility and $N$ the particle number.

After switching from the $(n,T)$ to the $(\Gamma,\kappa)$ state variables and expanding the derivatives, the Lebowitz-Percus correction for the YOCP becomes
\begin{align*}
&g_{\mathrm{c}}(r;\Gamma,\kappa)=\left[1+\frac{\chi_{\mathrm{T}}}{N}\right]g_{\mathrm{MD}}(r;\Gamma,\kappa)+\frac{\chi_{\mathrm{T}}}{2N}\left\{\frac{\Gamma^2}{9}\frac{\partial^2}{\partial{\Gamma}^2}+\right.\\&\,\,\left.\frac{\kappa^2}{9}\frac{\partial^2}{\partial{\kappa}^2}-\frac{2\Gamma\kappa}{9}\frac{\partial^2}{\partial\Gamma\partial\kappa}+\frac{10\Gamma}{9}\frac{\partial}{\partial\Gamma}-\frac{8\kappa}{9}\frac{\partial}{\partial\kappa}\right\}g_{\mathrm{MD}}(r;\Gamma,\kappa)\,.
\end{align*}
The final term involves all possible first and second order partial derivatives of the radial distribution function with respect to the state variables. In principle, these can be evaluated with simulations by employing finite difference approximations. Nevertheless, the standard second-order central-difference scheme would require nine MD simulations for any state point which correspond to all the first neighbouring grid points of a two-dimensional computational stencil centered at the state point of interest. To avoid this formidable task, the relative importance of the final term has been assessed with the aid of IEMHNC-generated radial distribution functions. For all $16$ state points of interest, the results revealed that the final term has no impact on the computed bridge functions. Hence, a simplified version of the Lebowitz-Percus correction can be safely employed that reads as\,\cite{PollOCPB,sizeexp3}
\begin{equation}
g_{\mathrm{c}}(r;\Gamma,\kappa)=\left[1+\frac{\chi_{\mathrm{T}}}{N}\right]g_{\mathrm{MD}}(r;\Gamma,\kappa)\,.\label{LebowitzPercus}
\end{equation}
We note that discarding of the final term is equivalent to stating that the radial distribution function is relatively insensitive to very small density variations; a reasonable assumption above the Fisher-Widom line and especially close to crystallization.

The only obstacle remaining in the implementation of the correction for the explicit size effect has to do with the calculation of the isothermal compressibility. In absence of an established YOCP equation of state that remains very accurate in the entire liquid domain, two alternative paths were followed. \emph{(a) Integral equation theory}. Given the well-documented successes of the IEMHNC approach for YOCP liquids, IEMHNC-produced radial distribution functions were employed. The IEMHNC approximation was solved in very dense extended grids of $\Delta{r}/d=0.0001$ and $\max{(r/d)}=100$. The virial route as well as the statistical route were considered. In the virial route to $\chi_{\mathrm{T}}$, the first derivatives of the pressure with respect to $(\Gamma,\kappa)$ were computed with the second-order central difference scheme and $\Delta\Gamma=\Delta\kappa=0.001$ finite steps. In the statistical route to $\chi_{\mathrm{T}}$, the integral relation that contains the direct correlation function was preferred and the asymptotic limit was added and then subtracted to reduce the truncation errors\,\cite{ourwork1}. \emph{(b) Hypervirial route}. This path is only realizable in MD simulations, since the ensemble averaged hypervirial is a non-thermodynamic quantity. The hypervirial relation reads as\,\cite{sizeexp4,sizeexp5}
\begin{equation}
\langle(\delta\mathcal{W})^2\rangle=N-\frac{N}{\chi_{\mathrm{T}}}+\langle\mathcal{W}\rangle+\langle\mathcal{X}\rangle\,.\label{HyperVirial}
\end{equation}
In the above; the dimensionless virial function is defined by $\mathcal{W}=-(1/3)\sum_{i}\sum_{j>i}\beta{w}(r_{ij})$ with the interatomic virial given by $w(r)=rdu/dr$, the dimensionless hypervirial function is defined by $\mathcal{X}=(1/9)\sum_{i}\sum_{j>i}\beta{x}(r_{ij})$ with the interatomic hypervirial given by $x(r)=rdw/dr$, the operator $\langle...\rangle$ denotes canonical ensemble averaging and $\delta\mathcal{W}=\mathcal{W}-\langle\mathcal{W}\rangle$ for the canonical deviations from mean. The MD implementation only requires the recording of the virial and hypervirial quantities at each time step. For all the $16$ YOCP state points of interest, the $\chi_{\mathrm{T}}$ values stemming from the three routes are reported in Table \ref{compressibilitytable}. The MD-hypervirial route will be preferred, but it is worth pointing out that the IEMHNC-virial route provides very accurate results while the IEMHNC-statistical route typically leads to slight underestimations.

\begin{figure}
	\centering
	\includegraphics[width=3.35in]{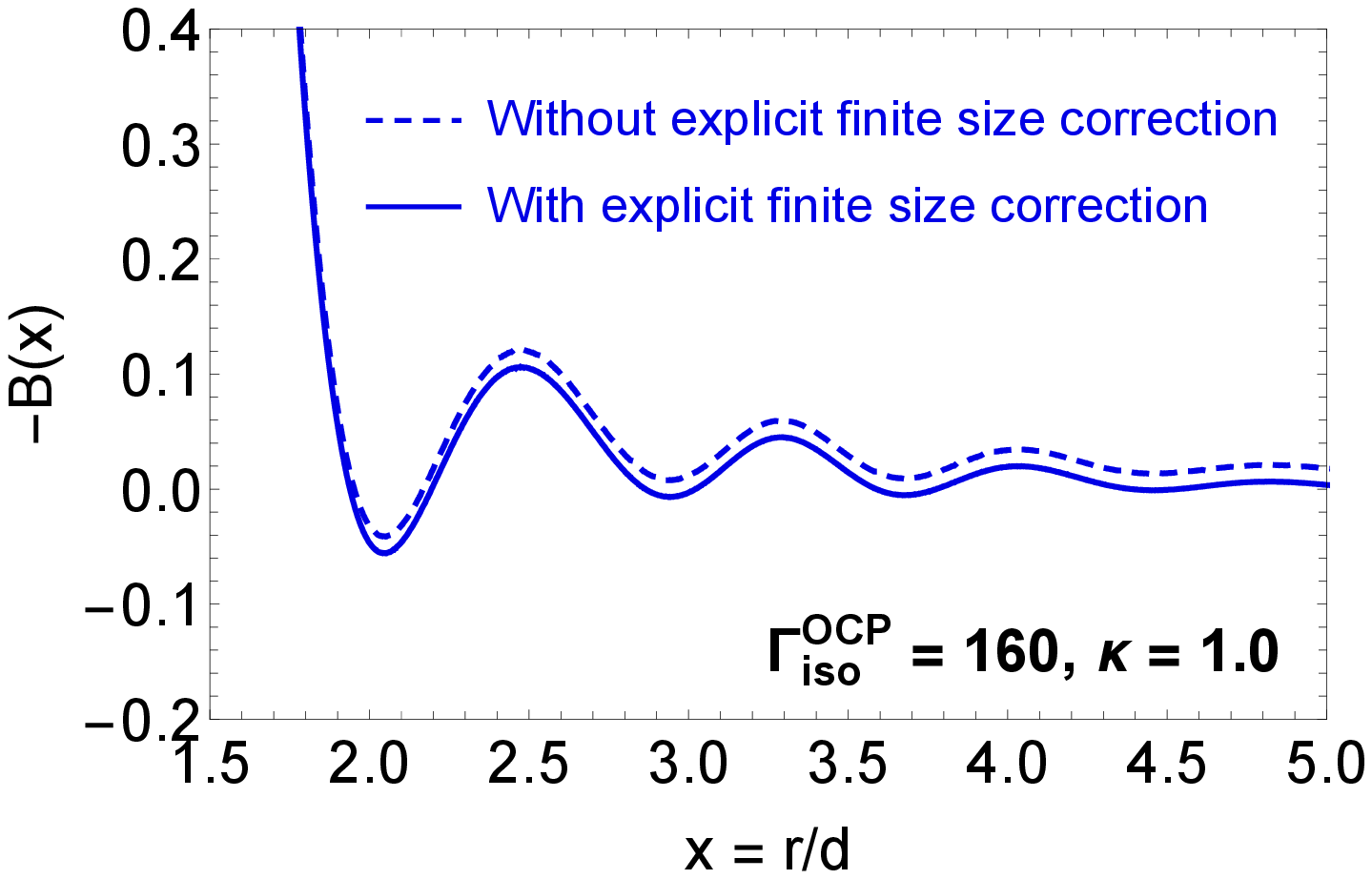}
	\caption{\emph{Finite size errors} in the computation of the YOCP bridge function at the state point $\Gamma_{\mathrm{ISO}}^{\mathrm{OCP}}=160,\,\kappa=1.0$. Results with and without the Lebowitz-Percus correction for the explicit finite size error.}\label{fig:size_error}
\end{figure}

\begin{table*}
	\caption{The reduced inverse isothermal compressibility $\mu_{\mathrm{T}}=1/\chi_{\mathrm{T}}$ for all the $16$ YOCP state points of interest. Results from the MD-hypervirial route (superscript MD), the IEMHNC-virial route (superscript IEMHNC,v) and the IEMHNC-statistical route (superscript IEMHNC,s). The standard deviations of the mean values of $\mu_{\mathrm{T}}^{\mathrm{MD}}$ are denoted by $e_{\mathrm{T}}^{\mathrm{MD}}$, the absolute relative deviations between the hypervirial and virial results are denoted by $\epsilon_{\mathrm{IEMHNC,v}}$ and the absolute relative deviations between the hypervirial and statistical results are denoted by $\epsilon_{\mathrm{IEMHNC,s}}$. The MD-hypervirial results are extremely accurate as inferred from the smallness of $e_{\mathrm{T}}^{\mathrm{MD}}/\mu_{\mathrm{T}}^{\mathrm{MD}}$ and have been preferred. The IEMHNC-virial results are also very accurate with relative deviations that are always less than $0.17\%$, while the IEMHNC-statistical results are less accurate with relative deviations ranging from $0.3\%$ up to $8.2\%$. This is expected given the small degree of thermodynamic inconsistency of the IEMHNC approach\,\cite{ourwork1,ourwork3}.}\label{compressibilitytable}
	\centering
	\begin{tabular}{ccccccccc}\hline
$\Gamma_{\mathrm{ISO}}^{\mathrm{OCP}}$ & $\kappa$                & $\Gamma$ & $\mu_{\mathrm{T}}^{\mathrm{MD}}$ & $e_{\mathrm{T}}^{\mathrm{MD}}$ & $\mu_{\mathrm{T}}^{\mathrm{IEMHNC,v}}$ & $\epsilon_{\mathrm{IEMHNC,v}}$ & $\mu_{\mathrm{T}}^{\mathrm{IEMHNC,s}}$ & $\epsilon_{\mathrm{IEMHNC,s}}$ \\ \hline
160\,\,\,\,\,                          & \,\,\,\,\,1.0\,\,\,\,\, &  205.061 & \,\,\,\,\,540.296\,\,\,\,\,      &  0.009                         & 540.252                                & 0.008\%                        & 531.436                                & 1.640\%                        \\
160\,\,\,\,\,                          & \,\,\,\,\,1.5\,\,\,\,\, &  286.437 & \,\,\,\,\,284.173\,\,\,\,\,      &  0.010                         & 284.649                                & 0.168\%                        & 274.49                                 & 3.407\%                        \\
160\,\,\,\,\,	                       & \,\,\,\,\,2.0\,\,\,\,\, &  435.572 & \,\,\,\,\,193.548\,\,\,\,\,      &  0.032                         & 193.478                                & 0.036\%                        & 182.123                                & 5.903\%                        \\
160\,\,\,\,\,	                       & \,\,\,\,\,2.5\,\,\,\,\, &  708.517 & \,\,\,\,\,150.459\,\,\,\,\,      &  0.062                         & 150.346                                & 0.075\%                        & 138.158                                & 8.176\%                        \\
120\,\,\,\,\,	                       & \,\,\,\,\,1.0\,\,\,\,\, &  153.796 & \,\,\,\,\,405.692\,\,\,\,\,      &  0.009                         & 405.674                                & 0.004\%                        & 400.909                                & 1.179\%                        \\
120\,\,\,\,\,	                       & \,\,\,\,\,1.5\,\,\,\,\, &  215.930 & \,\,\,\,\,215.176\,\,\,\,\,      &  0.017                         & 215.149                                & 0.013\%                        & 209.393                                & 2.688\%                        \\
120\,\,\,\,\,	                       & \,\,\,\,\,2.0\,\,\,\,\, &  328.816 & \,\,\,\,\,146.766\,\,\,\,\,      &  0.020                         & 146.757                                & 0.006\%                        & 140.145                                & 4.511\%                        \\
120\,\,\,\,\,	                       & \,\,\,\,\,2.5\,\,\,\,\, &  534.722 & \,\,\,\,\,114.437\,\,\,\,\,      &  0.045                         & 114.349                                & 0.077\%                        & 107.152                                & 6.366\%                        \\
80\,\,\,\,\,	                       & \,\,\,\,\,1.0\,\,\,\,\, &  102.531 & \,\,\,\,\,271.054\,\,\,\,\,      &  0.008                         & 271.056                                & 0.001\%                        & 269.65                                 & 0.518\%                        \\
80\,\,\,\,\,	                       & \,\,\,\,\,1.5\,\,\,\,\, &  144.330 & \,\,\,\,\,144.489\,\,\,\,\,      &  0.014                         & 144.517                                & 0.019\%                        & 142.533                                & 1.354\%                        \\
80\,\,\,\,\,	                       & \,\,\,\,\,2.0\,\,\,\,\, &  219.972 & \,\,\,\,\,99.007\,\,\,\,\,       &  0.024                         & 99.045                                 & 0.038\%                        & 96.553                                 & 2.479\%                        \\
80\,\,\,\,\,	                       & \,\,\,\,\,2.5\,\,\,\,\, &  357.136 & \,\,\,\,\,77.391\,\,\,\,\,       &  0.034                         & 77.457                                 & 0.085\%                        & 74.623                                 & 3.577\%                        \\
40\,\,\,\,\,	                       & \,\,\,\,\,1.0\,\,\,\,\, &  51.265  & \,\,\,\,\,136.347\,\,\,\,\,      &  0.006                         & 136.352                                & 0.004\%                        & 136.942                                & 0.436\%                        \\
40\,\,\,\,\,	                       & \,\,\,\,\,1.5\,\,\,\,\, &  72.537  & \,\,\,\,\,73.552\,\,\,\,\,       &  0.014                         & 73.577                                 & 0.038\%                        & 73.986                                 & 0.590\%                        \\
40\,\,\,\,\,	                       & \,\,\,\,\,2.0\,\,\,\,\, &  110.707 & \,\,\,\,\,50.936\,\,\,\,\,       &  0.018                         & 50.975                                 & 0.067\%                        & 51.206                                 & 0.530\%                        \\
40\,\,\,\,\,                           & \,\,\,\,\,2.5\,\,\,\,\, &  178.269 & \,\,\,\,\,40.021\,\,\,\,\,       &  0.022                         & 40.040                                 & 0.047\%                        & 40.156                                 & 0.337\%                        \\ \hline
	\end{tabular}
\end{table*}

Given its aperiodic form, see Eq.(\ref{LebowitzPercus}), the explicit finite size error can be expected to weakly affect the computed bridge functions despite the large simulated particle number ($N=54872$). The magnitude of this size error is controlled by the isothermal compressibility $\chi_{\mathrm{T}}$, whose thermodynamic state dependence is opposite to that of the bridge function sensitivity. In particular:\,(a) As $\Gamma/\Gamma_{\mathrm{m}}$ increases for constant screening parameter $\kappa$, $\chi_{\mathrm{T}}$ decreases but the sensitivity increases. (b) As $\kappa$ increases for a constant normalized coupling parameter $\Gamma/\Gamma_{\mathrm{m}}$, $\chi_{\mathrm{T}}$ increases but the sensitivity decreases. In both cases, the sensitivity variations are far more dramatic than the compressibility variations, which implies that the bridge functions of the state points that lie closer to the crystallization line and closer to the OCP limit should be more susceptible to explicit finite size errors.

The numerical results confirmed our theoretical expectations. The explicit finite size errors only influence the bridge functions along the isomorph line that is closer to the melting line and mainly affect the $\kappa=1$ member. The effect is minor and mostly manifested as a correction of the asymptotic behavior of the bridge function which now more properly converges to zero. A characteristic example is illustrated in figure \ref{fig:size_error}.

\subsubsection{Implicit finite size errors}\label{subErrorSizeImplicit}

\noindent The implicit size effect in computer simulations is a direct consequence of the spurious correlations which are introduced by imposing periodic boundary conditions\,\cite{sizeimp1,sizeimp2,sizeimp3}. In simple liquids that are simulated with standard cubic boxes, the infinite system should be viewed as a primitive cubic crystal which results from periodic repetition of the complex composite unit cell in all directions\,\cite{sizeimp2}. Such a picture demonstrates that the imposed symmetry of the periodic boundary is equivalent to a progressively weaker rigid bond between each unit cell particle and its infinite periodic images, which induces orientational order in the liquid. This leads to slightly non-isotropic radial distribution functions and to systematic errors in the spherically averaged radial distribution functions\,\cite{sizeimp4}.

Within the grand-canonical ensemble and with the aid of cluster expansion techniques, Pratt and Haan devised a formally exact theory for the implicit finite size effect under the assumption that particles do not directly interact with any of their periodic images\,\cite{sizeimp2}, which is valid for pair interaction potentials that are truncated within the unit cell. When neglecting a class of bridged graphs, an approximate expression was derived that connects the slightly anisotropic (explicit effect corrected) $g_{\mathrm{c}}(\boldsymbol{r})$ of the simulated system with the actual $g(r)$ of the bulk system. The expression has the superposition form and reads as
\begin{equation*}
g_{\mathrm{c}}(\boldsymbol{r}_{12})\simeq{g}(r_{12})\prod_{i,\mathrm{all}}g(\left|\boldsymbol{r}_1-\boldsymbol{r}_{2i}\right|)\,,
\end{equation*}
where $\boldsymbol{r}_{1}-\boldsymbol{r}_{2i}$ refers to the displacement vector between the particle $1$ and any of particle's $2$ infinite periodic images\,\cite{sizeimp2}. It is apparent that the above expression cannot be inverted with respect to the unknown bulk $g(r)$. In contrast to the explicit size effects, the implicit size effects cannot be directly corrected. However, under some reasonable additional assumptions, it has been shown that implicit size errors for the spherically averaged radial distribution function $g_{\mathrm{c}}(r)=(1/2\pi)\int{g}_{\mathrm{c}}(\boldsymbol{r})d\Omega$ scale as
\begin{equation*}
\frac{g(r)}{g_{\mathrm{c}}(r)}\sim1+\mathcal{O}\left[\left(\frac{1}{N}\right)^{\nu/3}\right]\,.
\end{equation*}
for inverse-power law (IPL) potentials of $\nu-$order\,\cite{sizeimp3}. On the other hand, see Eq.(\ref{LebowitzPercus}), explicit size errors scale as $g_{\mathrm{c}}(r)/g_{\mathrm{MD}}(r)\sim1+\mathcal{O}\left[1/N\right]$. This implies that, for sufficiently short-ranged potentials $\nu>3$, implicit size errors should decrease much faster with the number of particles. Since in our simulations the explicit size effect was very weak, the implicit size effect should be expected to be insignificant for the $\kappa\geq1$ Yukawa potentials considered, which have shorter range than the $\nu=3$ IPL potential. In addition, implicit size errors are not aperiodic like the explicit size errors, but display a $\lambda_{\mathrm{\epsilon}}\simeq2d$ periodicity that is comparable to the distance between successive coordination cells\,\cite{sizeimp3}. As demonstrated earlier, the sensitivity of the bridge function to such radial distribution function errors is much smaller. Hence, it can be safely concluded that implicit size errors have negligible effect on the computed bridge functions.

\subsection{Tail errors}\label{subErrorTail}

\noindent Tail errors emerge from the finite extent of the primary cell of the simulations. These errors are generated by the fact that radial distribution functions can only be reliably extracted inside a restricted domain with the maximum distance naturally equal to half the length of the cubic box. Unless MD-extracted radial distribution functions have truly reached their asymptotic limit of unity prior to $r=L/2$, their direct implementation will lead to errors in the direct correlation functions that will magnify while propagating to the bridge functions. Even for MD simulation domains of moderate size, tail errors can be nearly eliminated with a long-range extrapolation method.

There are three common types of long-range extrapolation methods. \emph{(a) Verlet methods.} These techniques are based on integral equation theory. They extend the $g(r)$ simulation data by assuming the beyond $L/2$ validity of a traditional closure equation that follows the unique functionality condition $B[\gamma]=B[\gamma(r)]$. The Percus-Yevick closure $c(r)=g(r)\{1-\exp{[+\beta{u}(r)]}\}$\,\cite{tailtec1}, hypernetted chain closure $c(r)=g(r)-1-\beta{u}(r)-\ln{[g(r)]}$\,\cite{tailtec2,tailtec3} and mean spherical approximation closure $c(r)=-\beta{u}(r)$\,\cite{tailtec4} are typically invoked. The extrapolated $g(r)$ is obtained by numerically solving the OZ equation that is complemented with mixed closure conditions. \emph{(b) Baxter methods.} These techniques are also based on integral equation theory and assume the long-range validity of a traditional closure equation, but they consider the equivalent Baxter system of equations that emerges from the Wiener-Hopf factorization of the OZ equation\,\cite{tailtec5,tailtec6}. \emph{(c) Asymptotic methods.} Such techniques conjecture that the long-range (near-asymptotic) $rh(r)$ can be approximated by a discrete sum of exponentially decaying oscillating functions. The solution of the OZ equation or the Baxter equations is circumvented. The unknown algebraic parameters are chosen to fit the simulation results in a subinterval of the intermediate range that extends up to $r=L/2$\,\cite{tailtec2,tailtec7}.

The $N=54872$ particles employed in the present MD simulations correspond to $L/2\simeq30d$, which was considered long enough for $g(r)$ to be sufficiently close to unity so that the influence of tail errors on the bridge functions is completely negligible. This expectation was verified by employing the Verlet method and the asymptotic method for all the $16$ YOCP state points of interest.

\begin{figure}
	\centering
	\includegraphics[width=3.35in]{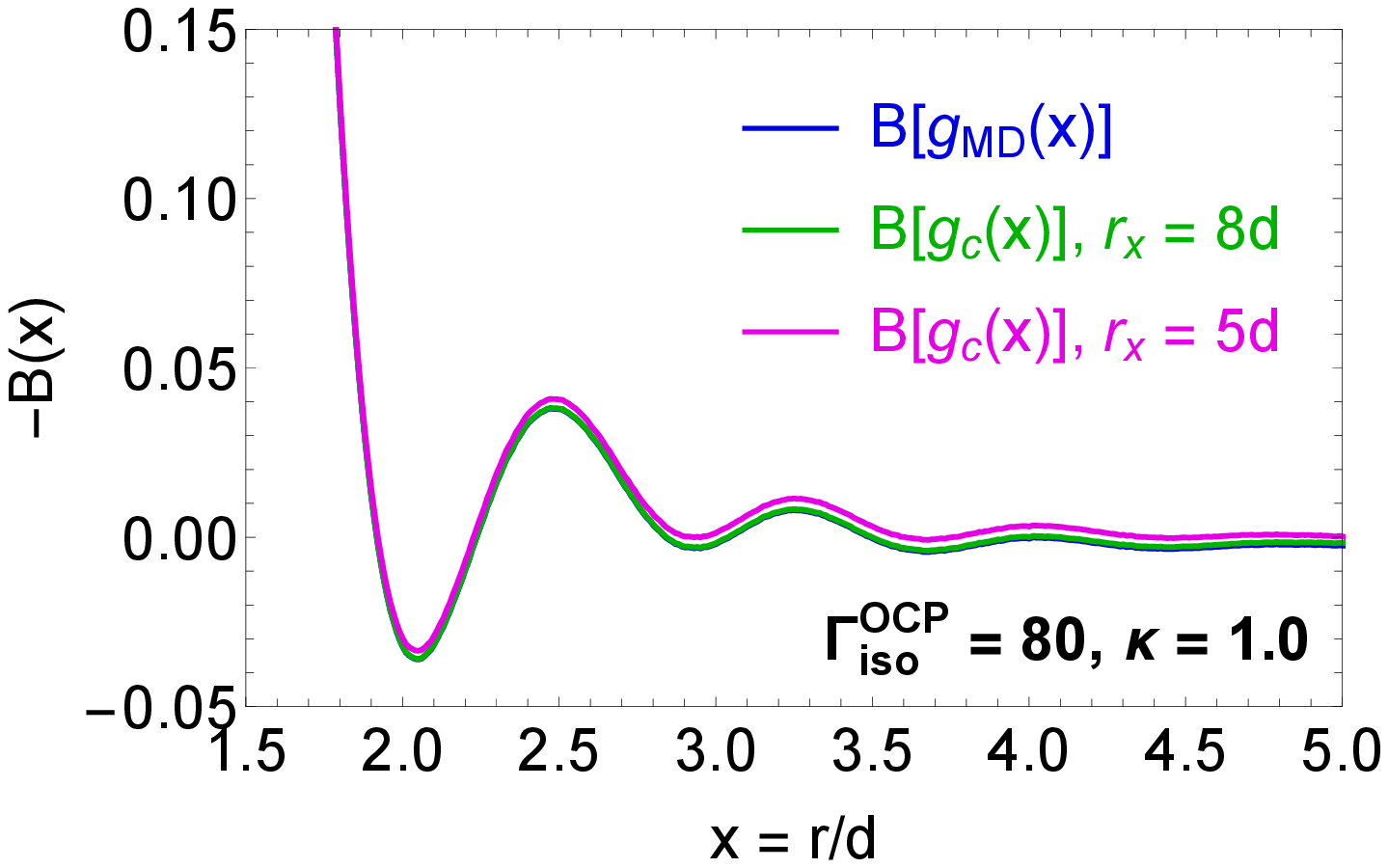}
	\caption{\emph{Tail errors} in the computation of the YOCP bridge function at the state point $\Gamma_{\mathrm{ISO}}^{\mathrm{OCP}}=80,\kappa=1.0$. Results for the extended radial distribution function extraction up to $30d$ without compensation for the tail errors and two limited radial distribution function extractions ($r_{\mathrm{x}}/d=5,\,8$) combined with the Verlet HNC extrapolation method. Tail errors can be totally compensated for $g(r)$ extractions at least up to $8d$, as seen by the indistinguishable respective bridge functions.}\label{fig:tail_error}
\end{figure}

The application of the asymptotic method was based on the assumption that $rh(r)$ can be approximated by a single exponentially decaying oscillatory term. The four unknown parameters $\{a_0,\,a_1,\,a_2,\,a_3\}$ that emerge in the expression $g_{\mathrm{fit}}(r)=1+(1/r)\exp{(a_0-a_1r)}\cos{(a_2+a_3r)}$ were determined by least square fitting to the extracted $g_{\mathrm{MD}}(r)$ in an interval $r_{\mathrm{f,d}}<r<r_{\mathrm{f,u}}$. The upper fitting limit was always retained at $r_{\mathrm{f,u}}=L/2\simeq30d$, while the lower fitting limit varied within $r_{\mathrm{f,d}}=15-10d$ depending on the state point. The negligible fitting errors supported the above functional assumption for $g_{\mathrm{fit}}(r)$. The extrapolated radial distribution function reads as
\begin{equation*}
 g_{\mathrm{c}}(r)=
    \begin{cases}
    g_{\mathrm{MD}}(r),\,\,\, 0\leq{r}\leq{r}_{\mathrm{x}}\qquad\qquad\qquad\qquad\qquad\,\,\,  \\
    g_{\mathrm{fit}}(r),\,\,\,\,\,\,{r}_{\mathrm{x}}\leq{r}\leq{r}_{\mathrm{m}}
    \end{cases}
\end{equation*}
where $r_{\mathrm{m}}/d=60$ was selected in order to double the numerical extraction domain and where ${r}_{\mathrm{x}}\in[r_{\mathrm{f,d}},r_{\mathrm{f,u}}]$ was selected so that the discontinuity between $g_{\mathrm{MD}}(r),g_{\mathrm{fit}}(r)$ is minimized (${r}_{\mathrm{x}}/d=13.6-28.3$ depending on the state point). The bridge functions that result from $g_{\mathrm{MD}}(r)$ up to $r=L/2\simeq30d$ and also from $g_{\mathrm{c}}(r)$ up to $r=60d$ were computed, corrected for finite size errors and compared. For all YOCP state points, the bridge functions totally overlapped, confirming that tail errors are negligible.

The Verlet method was utilized in combination with the hyper-netted chain (HNC) closure. The OZ equation has been supplemented with the closure condition
\begin{equation*}
 g_{\mathrm{c}}(r)=
    \begin{cases}
    g_{\mathrm{MD}}(r),\qquad\qquad\qquad\qquad\qquad\,\,\, 0\leq{r}\leq{r}_{\mathrm{x}}  \\
    \exp{[-\beta{u}(r)+g_{\mathrm{c}}(r)-1-c(r)]},\,{r}_{\mathrm{x}}\leq{r}\leq{r}_{\mathrm{m}}
    \end{cases}
\end{equation*}
where we selected $r_{\mathrm{m}}/d=60$ in order to double the numerical extraction domain and where we probed different $r_{\mathrm{x}}/d=\{5,8,10,15,20\}$ values in order to investigate how limited the MD extraction domain should be for the tail errors to be impossible to correct by long range extrapolations. The system of the OZ integral equation and the non-linear closure condition has been solved with Picard iterations in Fourier space; the real space resolution was $\Delta{r}/d=0.002$ that is equal to the optimum bin width employed to extract $g_{\mathrm{MD}}(r)$, the reciprocal space resolution was $\Delta{q}=\pi{d}/r_{\mathrm{m}}$ and at each iteration $n$ convergence was assumed when the criterion $||c_{n}(r)-c_{n-1}(r)||<10^{-8}$ was satisfied for all distances. The bridge functions that result from $g_{\mathrm{MD}}(r)$ up to $r=L/2\simeq30d$ and also from $g_{\mathrm{c}}(r)$ up to $r=60d$ were computed, corrected for finite size errors and compared. For all the YOCP state points, the bridge functions totally overlapped at least for $r_{\mathrm{x}}/d\geq8$ which demonstrates that tail errors are negligible for our simulations where $r_{\mathrm{x}}/d=30$. Even for $r_{\mathrm{x}}/d=5$, the deviations were very small along the $\Gamma_{\mathrm{ISO}}^{\mathrm{OCP}}=160$ isomorph and remained negligible along the $\Gamma_{\mathrm{ISO}}^{\mathrm{OCP}}=40$ isomorph, see figure \ref{fig:tail_error} for an illustration. The results suggest that the radial distribution functions did not need to be extracted up to $30d$; an extraction up to $8d$ combined with a reliable extrapolation method would have led to identical bridge functions with far less computational cost.

\subsection{Isomorphic errors}\label{subErrorIsomorph}

\noindent Isomorphic state point uncertainties emerge due to the fact that isomorphic curves are not traced out exactly. To be specific, the direct isomorph check is based on an approximate expression and it requires input from NVT\,MD simulations, whereas the small step method is based on an exact differential equation that is solved numerically and it also requires input from NVT\,MD simulations. As isomorphic curves are being traced out, both methods involve a controlled density re-scaling. As a consequence, densities are known exactly and only temperature uncertainties exist. For the YOCP, this implies that there are only uncertainties in the coupling parameters of the isomorphic state points. The isomorphic errors emerge from the propagation of these uncertainties to the bridge function. There is a direct propagation through the presence of the pair interaction potential and an indirect propagation through the presence of the radial distribution and direct correlation functions in the closure equation.

\begin{figure}
	\centering
	\includegraphics[width=3.25in]{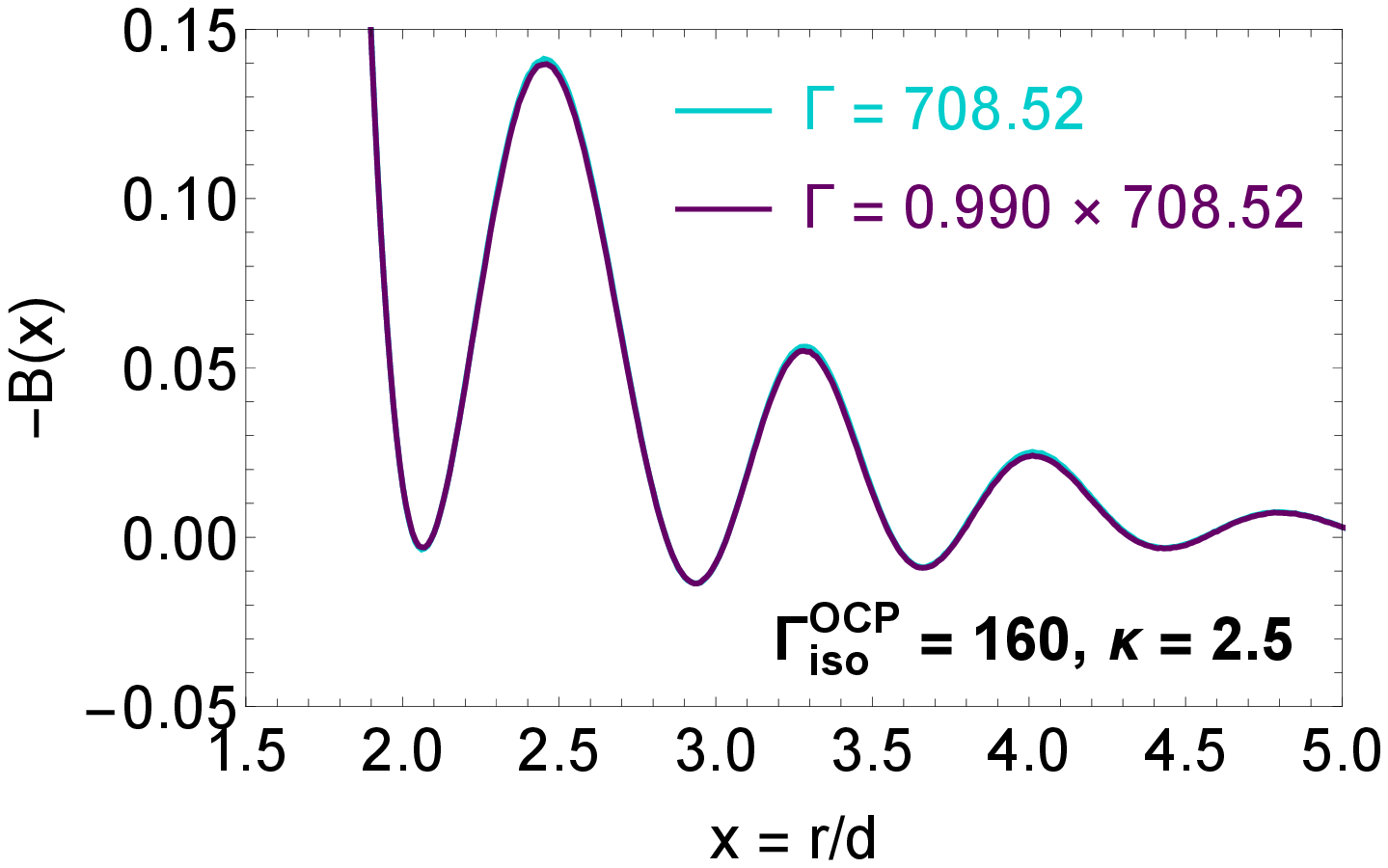}
	\caption{\emph{Isomorphic errors} in the computation of the YOCP bridge function at the state point $\Gamma_{\mathrm{ISO}}^{\mathrm{OCP}}=160,\kappa=2.5$. The errors are generated by the maximum $\Delta\Gamma/\Gamma=1\%$ isomorphic coupling parameter uncertainty. The results for $B(r;\Gamma,\kappa)$
    and $B(r;\Gamma-\Delta\Gamma,\kappa)$ are indistinguishable.}\label{fig:isomorphic_error1}
\end{figure}

\begin{figure}
	\centering
	\includegraphics[width=3.25in]{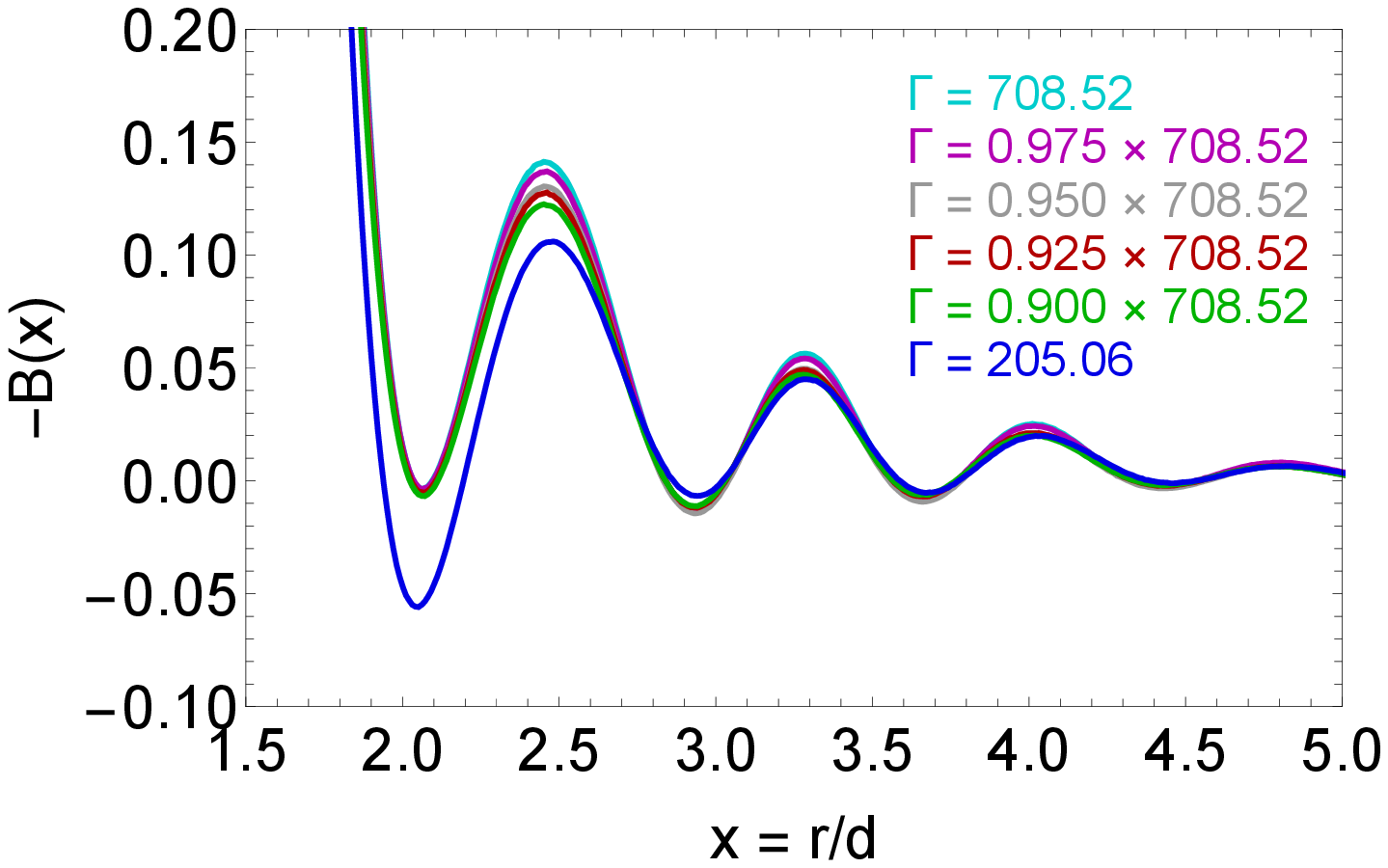}
	\caption{\emph{Isomorphic errors} in the computation of the YOCP bridge function at the state point $\Gamma_{\mathrm{ISO}}^{\mathrm{OCP}}=160,\kappa=2.5$. The errors are generated by artificial $\Delta\Gamma/\Gamma=2.5,\,5.0,\,7.5,\,10.0\%$ uncertainties in the isomorphic coupling parameter. Results
    for $B(r;\Gamma,\kappa=2.5)$, $B(r;\Gamma-\Delta\Gamma,\kappa=2.5)$ and $B(r;\Gamma^{\prime},\kappa=1.0)$, \emph{i.e.} the $\kappa=1.0$ member of the $\Gamma_{\mathrm{ISO}}^{\mathrm{OCP}}=160$ isomorph. Even $10\%$ uncertainties do not suffice to make the bridge functions of the $\kappa=2.5$ and $\kappa=1.0$ members fully invariant.}\label{fig:isomorphic_error2}
\end{figure}

The isomorphic coupling parameter uncertainties have been quantified from analysis of the local truncation errors and the global accumulated errors of the fourth order Runge-Kutta method that also accounted for statistical errors in MD-extracted density scaling exponents. Isomorph tracings based on the small step method were carried out with different logarithmic density increments ($|\Delta{n}|/n=2.3\%,\,4.8\%,\,8.8\%$) and particle numbers ($N=8192,\,17576$) in order to verify the output of this analysis. The empirical deviations between the isomorphs that stem from the direct isomorph check and the small step method were also considered. The emerging upper uncertainty thresholds varied within $\Delta\Gamma/\Gamma<0.5\%-1.0\%$ with the exact values depending on the screening parameter.

Isomorphic errors can be quantified by extracting the radial distribution function $g(r;\Gamma\pm\Delta\Gamma,\kappa)$ from MD simulations, computing the bridge function $B(r;\Gamma\pm\Delta\Gamma,\kappa)$ and comparing with $B(r;\Gamma,\kappa)$. It is preferable that grid errors are minimized as well as that finite size errors and tail errors are corrected in both bridge functions prior to comparison. This was pursued for the four state points corresponding to the $\kappa=2.5$ members of all YOCP isomorphs. These state points are subject to the largest isomorphic coupling parameter uncertainties $\Delta\Gamma/\Gamma<0.01$, since the sequential tracing of the isomorphs with the direct isomorph check and the small step method initiates at $\kappa=1.0$. Comparison revealed that isomorphic errors are negligible, see figure \ref{fig:isomorphic_error1}. In particular, the two bridge functions were indistinguishable for $\Gamma_{\mathrm{ISO}}^{\mathrm{OCP}}=40,80,120$ and very minor deviations were observed for $\Gamma_{\mathrm{ISO}}^{\mathrm{OCP}}=160$ that were much smaller than the statistical errors.

A similar procedure was followed in an attempt to answer the closely-related question of how much isomorphic coupling parameter uncertainties would be required in order to make the bridge functions of different isomorphic state points fully invariant. This problem was pursued for the $\Gamma_{\mathrm{ISO}}^{\mathrm{OCP}}=160$ isomorph in the following manner. The coupling parameter of the $\kappa=2.5$ member was reduced by $2.5\%,\,5.0\%,\,7.5\%,\,10\%$, the respective radial distribution functions were extracted from MD simulations and the respective bridge functions were computed. The latter were then compared with the bridge functions of the $\kappa=1.0,\,1.5,\,2.0$ members of the same isomorph. Comparison revealed that $\Gamma$ uncertainties exceeding $10\%$ are required in order to make the $\kappa=2.5$ and $\kappa=1.0$ bridge functions invariant, see figure \ref{fig:isomorphic_error2}.

\section{Corrected bridge functions including uncertainties}\label{InversionErrors}

\noindent As demonstrated in section \ref{RealUncertainties} for the MD simulation parameters of the production runs; the histogram bin width is selected in a manner which guarantees that grid errors are much smaller than statistical errors, the nature of the Yukawa pair potential together with the large number of simulated particles ensure that implicit finite size errors are negligible, the extraction domain is extended enough so that tail errors are negligible, the accuracy of isomorph tracing methods is high enough so that isomorphic errors are negligible. Hence, in order to incorporate the uncertainties, it suffices to correct for explicit finite size errors and to include error bars that stem from statistical errors.

\begin{figure*}
	\centering
	\includegraphics[width=7.0in]{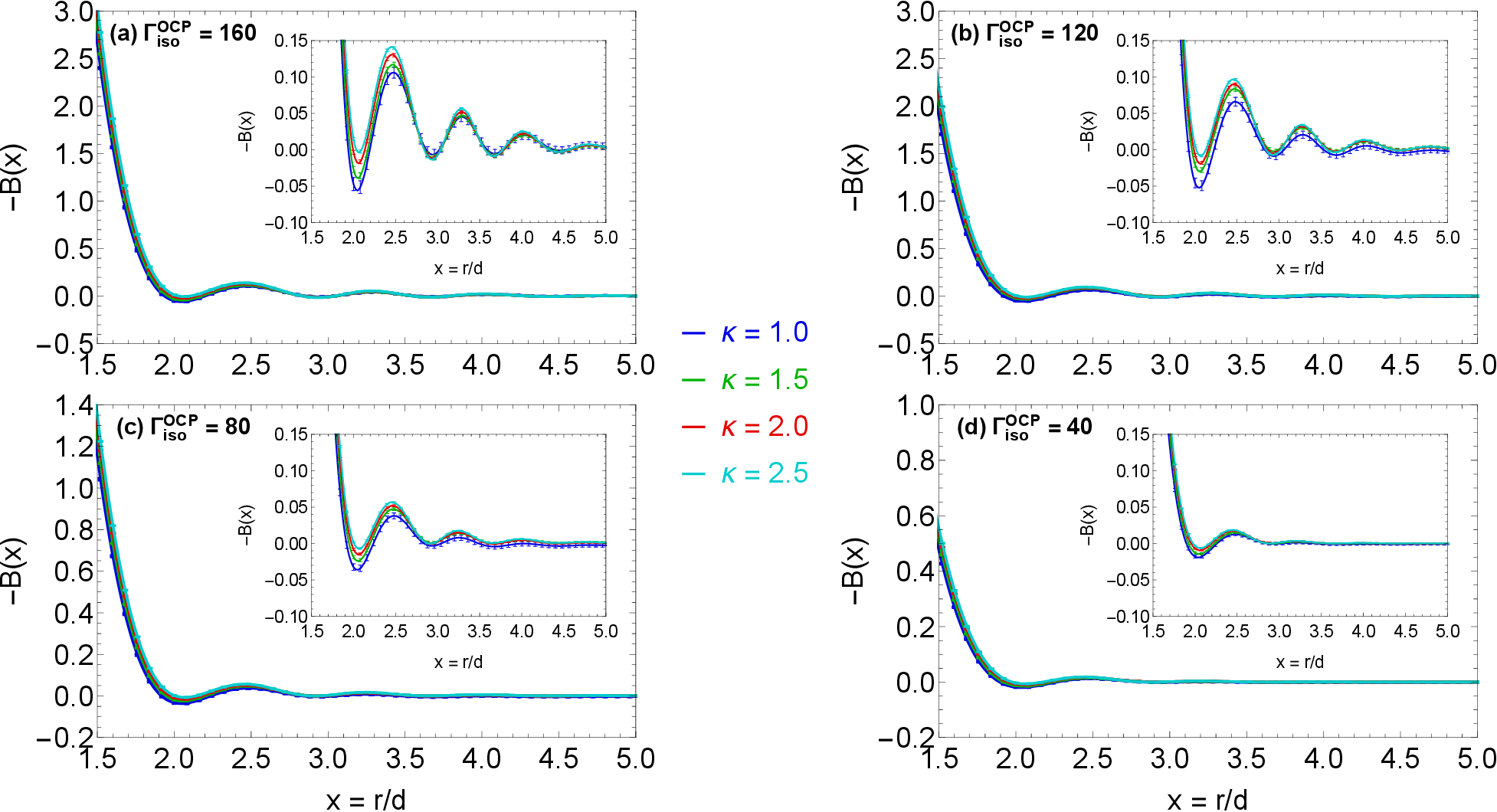}
	\caption{Finite-size corrected bridge functions featuring error bars due to the statistical uncertainties ($95\%$ confidence intervals) resulting from the application of the Ornstein-Zernike inversion method for dense YOCP liquids with radial distribution function input from NVT MD simulations (see the production runs). (a,\,b,\,c,\,d) Intermediate and long range results for the $4$ members of the $\Gamma_{\mathrm{ISO}}^{\mathrm{OCP}}=160,\,120,\,80,\,40$ isomorphs, respectively. The isomorphic deviations and the error bars are rather small and can be better discerned in the zoomed-in insets. The same applies for the quasi-periodic sign switching of the bridge functions.}\label{fig:invariance_error_bars}
\end{figure*}

The finite-size corrected bridge functions including statistical uncertainties are computed in the following manner for all the $16$ YOCP state points of interest. \textbf{\emph{First}}, the reduced isothermal compressibility is calculated from the production runs with the hypervirial route of Eq.(\ref{HyperVirial}). The simplified version of the Lebowitz-Percus correction, see Eq.(\ref{LebowitzPercus}), is now applied to the average radial distribution function $\langle{g}(r)\rangle_{M}$ and all block-averaged radial distribution functions $\langle{g}_i(r)\rangle_{N_{\mathrm{g}}}$. \textbf{\emph{Then}}, the average bridge function $B(r)=B[\langle{g}_{\mathrm{c}}(r)\rangle_{M}]$ and all the block-averaged bridge functions $B_i(r)=B[\langle{g}_{\mathrm{c},i}(r)\rangle_{N_{\mathrm{g}}}]$ are computed with the OZ inversion method from the respective size-corrected radial distribution functions. \textbf{\emph{Afterwards}}, the standard deviation of the average bridge function $\sigma[B(r)]$ is computed with the block averaging method outlined in section \ref{subErrorStatisticalBridge} by assuming the near optimal combination of $(N_{\mathrm{b}},\,N_{\mathrm{g}})=(256,\,256)$. \textbf{\emph{Finally}}, the standard error of the mean $\sigma[B(r)]$ is utilized for the determination of confidence intervals. In particular, the selected error bars for the statistical uncertainties correspond to $95\%$ confidence intervals.

The finite-size corrected bridge functions including error bars are illustrated in figure \ref{fig:invariance_error_bars} for the $4$ isomorphic curves and the $16$ YOCP state points of interest. It is evident that the errors cannot account for the small deviations that are observed between the bridge functions of the different members of the same isomorph. Therefore, the observed isomorph invariance of bridge functions in the long and intermediate range $r\geq1.5d$ is only approximate. Note that the relative invariance holds to nearly the same degree regardless of the YOCP isomorph. Note also that the approximate bridge function invariance extends up to the edges of the correlation void, where the approximate radial distribution function invariance begins to break down. Finally, it is worth pointing out that, for all the YOCP state points investigated, the bridge function becomes slightly positive close to $r\simeq2d$ well within the statistical uncertainties. In rough accordance with the asymptotic behavior of $B(r)=-(1/2)h^2(r)$ and with the $\sim1.5d$ periodicity of the total correlation function, additional shallower positive maxima appear with a periodicity slightly less than $\sim0.8d$. The emergence of sign switching and of a positive maximum within the first coordination cell seems to be a rather standard feature of the bridge functions of dense fluids that has also been observed for hard-sphere systems\,\cite{bridgHS1}, Lennard-Jones liquids\,\cite{bridgLJ1,bridgLJ2}, IPL-12 systems\,\cite{bridIPL1} and OCP liquids\,\cite{bridOCP1}. It is a salient feature of bridge functions that cannot be captured by the VMHNC approximation\,\cite{rosVMHNC} that utilizes the non-positive analytical Percus-Yevick hard-sphere bridge function. This deficiency has been suggested as responsible for the minor structural inaccuracies of the VMHNC approach that are observed in the vicinity of the first peak of the radial distribution function\,\cite{ourwork3}.

\section{Summary and conclusions}\label{Summary}

\noindent Yukawa bridge functions were systematically computed aiming to confirm or disprove the validity of the ansatz of reduced unit bridge function invariance along isomorphs, \emph{i.e.} phase diagram lines of constant excess entropy. $16$ state points were selected that belong to four isomorphs and cover the entire dense liquid YOCP phase diagram. The YOCP isomorphic curves were traced out with the small step method as well as the direct isomorph check. The intermediate and long bridge function ranges were made accessible after application of the Ornstein-Zernike inversion method with radial distribution function input from ultra-accurate molecular dynamics simulations that employed carefully selected parameters.

In order to accurately quantify the level of isomorph invariance of the bridge functions in the reliable extraction range, meticulous analysis of all sources of error was carried out. A detailed investigation of the sensitivity of bridge functions to aperiodic and periodic multiplicative perturbations in radial distribution functions led to important insights concerning the propagation of uncertainties. Regardless of the state point, it was consistently observed that YOCP bridge functions are far more sensitive to aperiodic perturbations. A strong phase diagram dependence was also discerned with the YOCP state points that lie in the vicinity of the melting line or near the OCP limit possessing more sensitive bridge functions. The use of these controlled artificial errors facilitated the understanding of all types of naturally emerging errors.

The statistical errors were quantified with a block averaging procedure that ultimately revealed a near-optimal combination of the block number and the sub-block configuration number after exhaustive trials. The grid errors were minimized compared to the omnipresent statistical errors due to the utilization of a near-optimal histogram bin width that emerged after comprehensive testing. The explicit finite size errors were observed to influence the bridge function asymptotes and were corrected with the simplified version of the Lebowitz-Percus expression valid for radial distribution functions that are relatively insensitive to small density variations, while the implicit finite size errors were deduced to be negligible after inspecting an approximate scaling derived from the exact Pratt and Haan rigid bond theory. The tail errors were confirmed to be negligible by the application of the Verlet and asymptotic long range extrapolation methods. The isomorphic errors that emerge from slight state point excess entropy mismatches due to inaccuracies in the isomorph tracing techniques were also shown to be negligible.

The final YOCP bridge functions, corrected for explicit finite size errors and featuring error bars stemming from the statistical uncertainties, were observed to be nearly isomorph invariant in the intermediate and the long range for all four excess entropies probed. This invariance was concluded to be approximate, since the small deviations observed between isomorphic bridge functions always exceed the quantified level of uncertainties. However, the bridge functions remain nearly isomorph invariant even at the edge of the correlation void, where the radial distribution functions and potentials of mean force already exhibit strong variance. The isomorph invariance of YOCP bridge functions well within the correlation void up to the origin is investigated in the accompanying paper\,\cite{accompan}.

\section*{Acknowledgments}

\noindent The authors would like to acknowledge the financial support of the Swedish National Space Agency under grant no.\,143/16. This work was also partially supported by VILLUM Foundation’s Grant No.\,16515 (Matter). GPU molecular dynamics simulations were carried out at the \emph{Glass and Time} computer cluster (Roskilde University). CPU molecular dynamics simulations were carried out on resources provided by the Swedish National Infrastructure for Computing (SNIC) at the NSC (Link{\"o}ping University) partially funded by the Swedish Research Council through grant agreement no.\,2016-07213.

\end{document}